\documentclass[
11pt,
prd
,preprint
,amsfonts,amsmath,amssymb
]{revtex4}
\usepackage{bm}
\usepackage[dvipdfmx]{graphicx}
\usepackage{color,float}
\begin{document}
\title{Reduced dynamics with Poincar\'{e} symmetry in an open quantum system}
\author{Akira Matsumura}
\email{matsumura.akira@phys.kyushu-u.ac.jp}
\affiliation{Department of Physics, Kyushu University, Fukuoka, 819-0395, Japan}
\begin{abstract}
We consider how the reduced dynamics of an open quantum system coupled to an environment admit the Poincar\'{e} symmetry.
The reduced dynamics are described by a dynamical map, which is a quantum channel (a completely positive and trace-preserving linear map) given by tracing out the environment from the total unitary evolution without initial correlations. 
We investigate the dynamical map invariant under the Poincar\'{e} transformations and discuss how the invariance constrains the form of the map.
Based on the unitary representation theory of the Poincar\'{e} group, 
we develop a systematic way to construct the dynamical map with the Poincar\'{e} invariance.  
Using this way, we derive such a dynamical map for a spinless massive particle, and the conservation of the Poincar\'{e} generators is discussed.
We then find the map with the Poincar\'{e} invariance and the four-momentum conservation. 
Further, we show that the conservation of the angular momentum and the boost operator makes the map of a spinless massive particle unitary.
\end{abstract}
\maketitle

\tableofcontents
\section{Introduction}

It is difficult to isolate a quantum system perfectly, which is affected by the inevitable influence of a surrounding environment. 
Such a quantum system is called an open quantum system.
Since we encounter open quantum systems in a wide range of fields such as quantum information science \cite{Breuer2016, Breuer2002}, condensed matter physics \cite{Feynman1963, Caldeira1983} and high energy physics \cite{Calzetta2008}, it is important to understand their dynamics.
In general, the dynamics of an open quantum system, the so-called reduced dynamics, are very complicated.
This is because the environment may have infinitely many uncontrollable degrees of freedom.
One needs the effective theory with relevant degrees of freedom to describe the reduced dynamics of an open quantum system \cite{Breuer2002}. 

As is well-known, symmetry gives a powerful tool for capturing relevant degrees of freedom in the dynamics of interest. 
For example, let us focus on the symmetry in the Minkowski spacetime, which is called the Poincar\'{e} symmetry. 
Imposing the Poincar\'{e} symmetry on a quantum theory, one finds that quantum dynamics in the theory are described by the fundamental degrees of freedom such as a massive particle and a massless particle \cite{Weinberg1995}.  
The approach based on symmetries may provide a way to get the effective theory of open quantum systems. 

In this paper, we discuss the consequences of the Poincar\'{e} symmetry on the reduced dynamics of an open quantum system. 
This is motivated for understanding relativistic theories of open quantum systems (for example, \cite{Bedingham2014,Bedingham2019, Jones2021a, Jones2021b, Pearle2015, Kurkov2011, Ahn2003, Wang2022, Meng2021}) and the theory of quantum gravity. 
At the present time, quantum mechanics and gravity have not been unified yet.
This situation has prompted to propose many models of gravitating quantum systems. 
In Ref.\cite{Kafri2014}, the model of a classical gravitational interaction between quantum systems was proposed, which is called the Kafri-Taylor-Milburn model. 
In addition, the Diosi-Penrose model \cite{Diosi1987,Diosi1989, Penrose1996} and the Tilloy-Diosi model \cite{Tilloy2016} were advocated, for which gravitating quantum system intrinsically decoheres. 
The above models are formulated in non-relativistic theories of open quantum systems. 
One may concern how they are incorporated in relativistic theories. 
This paper would help to obtain a relativistic extension of the above models. 

For our analysis, we describe the reduced dynamics of an open quantum system by a dynamical map. 
The dynamical map is a quantum channel obtained by tracing out the environment from the total unitary evolution with an initial product state. 
It is known that the dynamical map (the quantum channel) is a completely positive and trace-preserving linear map and has an operator-sum representation given by Kraus operators \cite{Breuer2002,Davies1976,Holevo2001,Keyl2002}. 
We consider the condition of a dynamical map invariant under the Poincar\'{e} transformations. 
It is first shown that this condition is satisfied for unitary evolution in quantum theory with the Poincar\'{e} symmetry. 
We then consider how the condition restricts the form of Kraus operators associated with the dynamical map.
With the help of the representation theory of the Poincar\'{e} group, we obtain a systematic way to deduce the Kraus operators. 

Applying this way, we get a model of the dynamical map of a spinless massive particle. 
Discussing the conservation of the Poincar\'{e} generators, we obtain the following consequences: (i) there is the non-unitary dynamical map (i.e. non-unitary channel) with the Poincar\'{e} invariance and the four-momentum conservation. 
(ii) If we impose the conservation of the Poincar\'{e} generators, then the map of a spinless massive particle is reduced to the unitary map (i.e. unitary channel) generated by the time-translation operator. 
These imply that the Poincar\'{e} symmetry can strongly constrain the reduced dynamics of an open quantum system. 
We further discuss a covariant formulation of the dynamical map with the Poincare invariance. 

The structure of this paper is as follows. 
In Sec. \ref{sec:2}, we discuss the dynamical map describing the reduced dynamics of an open quantum system and introduce the invariance of the dynamical map. 
In Sec. \ref{sec:3}, we derive the condition that the dynamical map is invariant under the Poincar\'{e} group. 
In Sec. \ref{sec:4}, focusing on the dynamics of a spinless massive particle, we present a model of the dynamical map with the Poincar\'{e} invariance.
We then investigate the model in terms of the conservation law of the Poincar\'{e} generators. 
Sec. \ref{sec:extra} is prepared for discussing a covariant formulation of our theory, and Sec. \ref{sec:5} is devoted as the conclusion. 
The natural unit
$\hbar=c=1$ is used in this paper. 

\section{Quantum dynamical map and its symmetry}
\label{sec:2}

In this section, we consider the reduced dynamics of an open quantum system and discuss the symmetry of the dynamics. 
The reduced dynamics are given as the time evolution of system density operator. 
The evolution from a time slice
$\tau=t_0$ to 
$\tau=t$ is
assumed to be given by 
\begin{equation}
\rho(t)=\Phi_{t, t_0}[\rho(t_0)]=\text{Tr}_\text{E}[\hat{U}(t,t_0) \rho(t_0) \otimes \rho_\text{E} \, \hat{U}^\dagger (t,t_0)],
\label{eq:Phi}
\end{equation}
where 
$\rho(\tau)$ is the system density operator,  
$\rho_\text{E}$ is the density operator of an environment, and 
$\hat{U}(t,t_0)$ 
is the unitary evolution operator of the total system. 
In this paper, the map 
$\Phi_{t, t_0}$ is called a dynamical map, which has the property called completely positive and trace-preserving \cite{Breuer2002, Davies1976, Holevo2001, Keyl2002}. 
The dynamical map 
$\Phi_{t,t_0}$ is rewritten in the operator-sum representation,
\begin{equation}
\Phi_{t,t_0}[\rho(t_0)]=\sum_{\lambda} \hat{F}^{t,t_0}_{\lambda} \, \rho(t_0) \, \hat{F}^{t,t_0 \,\dagger }_{\lambda} 
\label{eq:opsum},
\end{equation}
where  
$\hat{F}_{\lambda}^{t,t_0}$ called the Kraus operators. 
Note that this generally follows from Eq.\eqref{eq:Phi} (see, Appendix \ref{App:opsum}).
The Kraus operators satisfy the completeness condition,
\begin{equation}
\sum_{\lambda} \hat{F}^{t,t_0 \, \dagger}_{\lambda} \hat{F}_{\lambda}^{t,t_0}=\hat{\mathbb{I}}
\label{eq:FdgF}.
\end{equation}
which guarantees the trace-preserving property, 
$\text{Tr}[\Phi_{t,t_0}[\rho(t_0)]]=\text{Tr}[\rho(t_0)]$. 
In the operator-sum representation, 
$\lambda$ takes a discrete value. 
When 
$\lambda$ is a continuous value, we should replace the summation 
$\sum_\lambda$ with the integration
$\int d \mu(\lambda) $ with an appropriate measure 
$\mu(\lambda)$. 
It is known that two dynamical maps 
$\Phi$ and 
$\Phi'$ with 
\begin{equation}
\Phi[\rho]=\sum_{\lambda} \hat{F}_{\lambda} \, \rho \, \hat{F}^{\dagger }_{\lambda} , \quad 
\Phi'[\rho]=\sum_{\lambda} \hat{F}'_{\lambda} \, \rho \, \hat{F}^{'\dagger }_{\lambda}
\label{eq:PhiPhi},
\end{equation}
are equivalent to each other 
(i.e. $\Phi[\rho]=\Phi'[\rho]$ for any density operator 
$\rho$) if and only if there is a unitary matrix $\mathcal{U}_{\lambda \lambda'}$ 
satisfying $\sum_\lambda \mathcal{U}_{\lambda_1 \lambda} \mathcal{U}^*_{\lambda_2 \lambda}=\delta_{\lambda_1 \lambda_2}=\sum_\lambda \mathcal{U}_{\lambda \lambda_1} \mathcal{U}^*_{\lambda \lambda_2}$ and  
\begin{equation}
\hat{F}^{'}_\lambda=\sum_{\lambda'} \mathcal{U}_{\lambda \lambda'} \hat{F}_{\lambda'}.
\label{eq:inv}
\end{equation}
This is the uniqueness of a dynamical map \cite{Breuer2002, Davies1976, Holevo2001, Keyl2002}. 

To introduce symmetry in the above formulation, we schematically consider the differential equation of density operator, 
\begin{equation}
d\rho(\tau)  = d\mathcal{L}_{\tau} [\rho(\tau)],
\label{eq:drho}
\end{equation}
whose solution is 
$\rho(t)=\Phi_{t,t_0}[\rho(t_0)]$ from a time slice
$\tau=t_0$ to 
$\tau=t$. 
For example, if 
$d\mathcal{L}_\tau =d\tau \mathcal{L}$ with a Lindbladian 
$\mathcal{L}$, then 
$\Phi_{t,t_0}=e^{\mathcal{L}(t-t_0)}$, which is nothing but the superoperator of quantum dynamical semigroup \cite{Breuer2002, Davies1976, Holevo2001, Lindblad1976,Gorini1976}. 
The differential equation \eqref{eq:drho} is called covariant if the following equation, 
\begin{equation}
d\rho'(\tau) = d\mathcal{L}'_{\tau} [\rho'(\tau)],
\label{eq:drho'}
\end{equation}
holds under a transformation with 
$\rho(\tau) \rightarrow \rho'(\tau)$
and
$d\mathcal{L}_\tau \rightarrow d\mathcal{L}'_{\tau}$.
The map 
$d\mathcal{L}_\tau$ is invariant under the transformation when 
$d\mathcal{L}'_{\tau}=d\mathcal{L}_\tau$. 
This leads to the equation,
\begin{equation}
d\rho'(\tau) = d\mathcal{L}_\tau [\rho'(\tau)].
\label{eq:drho'=dPhi[rho']}
\end{equation}
We adopt the transformation rule given by 
$\rho'(\tau)=\hat{U}_{\tau}(g) \rho(\tau) \hat{U}^\dagger_{\tau} (g)$, where 
$\hat{U}_{\tau}(g)$ with 
$g \in G$ is the unitary representation of a group 
$G$.
Substituting 
$\rho'(\tau)=\hat{U}_{\tau}(g) \rho(\tau) \hat{U}^\dagger_{\tau} (g)$ into 
$d\rho'(\tau)=d\mathcal{L}_\tau [\rho'(\tau)]$ and solving the differential equation, 
we get
\begin{equation}
\hat{U}_{t}(g) \Phi_{t,t_0}[\rho(t_0)] \hat{U}^\dagger_{t} (g)=\Phi_{t,t_0}[\hat{U}_{t_0}(g) \rho(t_0) \hat{U}^\dagger_{t_0} (g)].
\label{eq:covmap}
\end{equation}
This defines that the dynamical map 
$\Phi_{t,t_0}$ is invariant under the group 
$G$, which was introduced in Refs.\cite{Keyl2002, Cirstoiu2020, Marvian2014}.
In the next section, we will see that Eq.\eqref{eq:covmap} holds for unitary evolution in quantum theory with the Poincar\'{e} symmetry. 
Our aim is to extend this to a general dynamical map and 
to construct the map which is invariant under the Poincar\'{e} group.

\section{Dynamical map with Poincar\'{e} invariance}
\label{sec:3}

In this section, we consider a quantum theory with the Poincar\'{e} symmetry and introduce the dynamical map with the Poincare invariance. 
The generators of the unitary representation of the Poincar\'{e} group \cite{Weinberg1995} are given by
\begin{equation}
\hat{P}^\mu =\int d^3 x \hat{T}^{0\mu}, \quad \hat{J}^{\mu \nu}=\int d^3x \hat{M}^{\mu \nu 0} ,
\label{eq:PJ}
\end{equation}
where 
$\hat{T}^{\mu \nu}$
is the energy-momentum tensor of a system,
and 
$\hat{M}^{\mu \nu \rho}$ is defined as
\begin{equation}
\hat{M}^{\mu \nu \rho}=x^\mu \hat{T}^{\nu \rho}-x^\nu \hat{T}^{\mu \rho}.
\label{eq:M}
\end{equation}
In the Schr\"{o}dinger picture, each component of the generators is
\begin{align}
&\hat{H}=\hat{P}^0=\int d^3 x \, \hat{T}^{00}(\bm{x},0),
\label{eq:H}
\\
&
\hat{P}^i=\int d^3 x \, \hat{T}^{0i}(\bm{x},0),
\label{eq:P}
\\
&\hat{J}^i=\frac{1}{2} \epsilon^{jki} \hat{J}_{jk} = \int d^3 x \, \epsilon^{jki}x_j \hat{T}^{0}_k (\bm{x},0), 
\label{eq:J}
\\
&\hat{K}^i (t) =\hat{J}^{i0}=\int d^3 x \, \big[x^i \hat{T}^{00}
(\bm{x},0)-t\hat{T}^{0i}(\bm{x},0) \big],
\label{eq:K}
\end{align}
where note that the boost generator 
$\hat{K}^i (t)$ explicitly depends on a time 
$t$. 
The operators satisfy the Poincar\'{e} algebra, 
\begin{align}
[\hat{P}_i, \hat{P}_j] &=0, 
\label{eq:PP}
\\
[\hat{P}_i, \hat{H}] &=0,
\label{eq:PH}
\\
[\hat{J}_i, \hat{H}] &=0,
\label{eq:JH}
\\
[\hat{J}_i, \hat{J}_j]&=i\epsilon_{ijk} \hat{J}^k,
\label{eq:JJ}
\\
[\hat{J}_i, \hat{P}_j]&=i\epsilon_{ijk} \hat{P}^k,
\label{eq:JP}
\\
[\hat{J}_i, \hat{K}_j]&=i\epsilon_{ijk} \hat{K}^k,
\label{eq:JK}
\\
[\hat{K}_i, \hat{P}_j]&=i \delta_{ij} \hat{H},
\label{eq:KP}
\\
[\hat{K}_i, \hat{H}]&=i \hat{P}_i,
\label{eq:KH}
\\
[\hat{K}_i, \hat{K}_j]&=-i \epsilon_{ijk}\hat{J}^k.
\label{eq:KK}
\end{align}

For later analysis, we clarify the explicit time dependence of the boost generator 
$\hat{\bm{K}} (t)=[\hat{K}^1(t),\hat{K}^2(t),\hat{K}^3(t)]$.
Since the boost operator is conserved during the evolution generated by the system Hamiltonian 
$\hat{H}$, we have the conservation law, 
$d\hat{\bm{K}}_\text{H}(t)/dt=0$, with 
$\hat{\bm{K}}_\text{H} (t)=e^{i\hat{H}t} \hat{\bm{K}}(t) e^{-i\hat{H}t}$. 
The solution of 
$d\hat{\bm{K}}_\text{H}(t)/dt=0$ is 
$\hat{\bm{K}}_\text{H}(t)=\hat{\bm{K}}_\text{H}(0)=\hat{\bm{K}}(0)$, and 
hence we get
\begin{equation}
\hat{\bm{K}} (t) =e^{-i\hat{H}t} \, \hat{\bm{K}} (0) \, e^{i\hat{H}t},
\label{eq:KK_0}
\end{equation}
where 
\begin{equation}
\hat{\bm{K}} (0)=\int d^3 x \, \bm{x} \hat{T}^{00} (\bm{x},0). 
\label{eq:K_0}
\end{equation}

Let us assume that the system dynamics are described by a dynamical map 
$\Phi_{t, t_0}$ from 
$\rho(t_0)$ to 
$\rho(t)=\Phi_{t, t_0}[\rho(t_0)]$, where 
$\rho(\tau)$ is the system density operator. 
According to Eq.\eqref{eq:covmap}, the Poincar\'{e} invariance of the dynamical map is formulated as
\begin{equation}
\hat{U}_{t} (\Lambda, a) \Phi_{t,t_0}[\rho(t_0)] \hat{U}^\dagger_{t} (\Lambda, a)=\Phi_{t,t_0}[\hat{U}_{t_0} (\Lambda, a)\rho(t_0)\hat{U}^\dagger_{t_0} (\Lambda, a)], 
\label{eq:P-Cov}
\end{equation}
where the unitary operator 
$\hat{U}_{t} (\Lambda, a)$ depends on the proper 
($\text{det} \Lambda =1$) orthochronous 
($\Lambda^0{}_0 \geq 1$) Lorentz transformation matrix 
$\Lambda^\mu {}_\nu$ and on
the real parameters 
$a^\mu$ of spacetime translations. 
The unitary operator 
$\hat{U}_{t} (\Lambda, a)$ 
generated by 
$\hat{H}, \hat{\bm{P}}=[\hat{P}^1,\hat{P}^2,\hat{P}^3], \hat{\bm{J}}=[\hat{J}^1,\hat{J}^2,\hat{J}^3]$ and 
$\hat{\bm{K}}(t)$ has the group multiplication rule 
\begin{equation}
\hat{U}_{t} (\Lambda', a') \hat{U}_{t} (\Lambda,a)= \hat{U}_{t} (\Lambda'\Lambda, a'+\Lambda' a), 
\label{eq:rule}
\end{equation}
where we adopted the non-projective unitary representation of the Poincar\'{e} group \cite{Weinberg1995}. 
The time dependence of 
$\hat{U}_t$ comes from the boost generator 
$\hat{\bm{K}}(t)$. 

Here, it is worth discussing the present approach and emphasizing the scope of this paper.
One may imagine that the reduced dynamics of a system are not Poincar\'{e} invariant even if the total dynamics of the system and its surrounding environment \eqref{eq:Phi} are Poincar\'{e} invariant. 
In the present analysis, we just investigate the form of the dynamical map satisfying \eqref{eq:P-Cov}, and we do not care about how such a dynamical map is derived from the total dynamics of the system and the environment. 
In general, the system reduced dynamics depend not only on the interaction Hamiltonian between the system and its surrounding environment, but also on the initial state of the environment. 
Hence, it is possible to find equivalent reduced dynamics obtained from different models of environment.
From this viewpoint, the approach based on a dynamical map (quantum channel) is independent of how the map is derived from possible models.
Nevertheless, it is important for discussing what model of environment gives the dynamical map with the Poincar\`{e} invariance.
This would help us to grasp a physical picture of the present approach.
In this paper, we do not investigate such a model, and the study on the model is listed as a future issue.

Before we start analyzing the dynamical map consistent with \eqref{eq:P-Cov}, let us understand how the condition \eqref{eq:P-Cov} holds for the unitary map, 
\begin{equation}
\mathcal{U}_{t,t_0}[\rho(t_0)]= e^{-i\hat{H}(t-t_0)} \, \rho(t_0) \, e^{i\hat{H}(t-t_0) }.
\label{eq:uts}
\end{equation}
According to the Poincar\'{e} algebra and Eq.\eqref{eq:KK_0}, we have 
\begin{equation}
\hat{U}_{t}(\Lambda,a)=e^{-i\hat{H}t} \hat{U}_0 (\Lambda, a) e^{i\hat{H}t},
\label{eq:U_t=0}
\end{equation}
where 
$\hat{U}_0 (\Lambda,a)$ is the unitary representation of the Poincar\'{e} group with the genrators 
$\hat{H}, \hat{\bm{P}}, \hat{\bm{J}}$ and 
$\hat{\bm{K}} (0)$. 
Using Eq.\eqref{eq:U_t=0}, we can check the invariance condition \eqref{eq:P-Cov} of the unitary map as
\begin{align}
\mathcal{U}_{t,t_0}[\hat{U}_{t_0} (\Lambda,a) \rho(t_0) \hat{U}^\dagger_{t_0} (\Lambda,a) ]
&= e^{-i\hat{H}(t-t_0)} \hat{U}_{t_0} (\Lambda,a)\, \rho(t_0) \, \hat{U}^\dagger_{t_0} (\Lambda,a)e^{i\hat{H}(t-t_0) }
\nonumber 
\\
&= e^{-i\hat{H}(t-t_0)} \hat{U}_{t_0} (\Lambda,a)e^{i\hat{H}(t-t_0)} \, \mathcal{U}_{t,t_0}[\rho(t_0)] \, e^{-i\hat{H}(t-t_0)}\hat{U}^\dagger_{t_0} (\Lambda,a)e^{i\hat{H}(t-t_0) }
\nonumber 
\\
&=  \hat{U}_t (\Lambda,a) \, \mathcal{U}_{t,t_0}[\rho(t_0)] \, \hat{U}^\dagger_t (\Lambda,a).
\nonumber 
\end{align}

Let us extend the invariant property of the unitary map to a general dynamical map. 
In the operator-sum representation, Eq.\eqref{eq:P-Cov} is written as
\begin{equation}
\hat{U}_{t} (\Lambda, a) \sum_{\lambda} \hat{F}^{t,t_0}_{\lambda} \, \rho(t_0) \, \hat{F}^{t,t_0 \dagger}_{\lambda} \, \hat{U}^\dagger_{t} (\Lambda, a)=\sum_{\lambda} \hat{F}^{t,t_0}_{\lambda} \, \hat{U}_{t_0} (\Lambda, a)\rho(t_0)\hat{U}^\dagger_{t_0} (\Lambda, a) \, \hat{F}^{t,t_0 \dagger }_{\lambda}.
\nonumber
\end{equation}
The uniqueness of the Kraus operators $\hat{F}^{t,t_0}_\lambda$ (see Eq.\eqref{eq:inv}) yields
\begin{equation}
\hat{U}^\dagger_{t} (\Lambda, a) \hat{F}^{t,t_0}_{\lambda} \hat{U}_{t_0} (\Lambda, a)= \sum_{\lambda'} \mathcal{U}_{\lambda \lambda'} (\Lambda,a) \hat{F}^{t,t_0}_{\lambda'}.
\label{eq:P-cov2}
\end{equation}
We can always choose 
$\hat{F}^{t,t_0}_\lambda$ so that 
$\{\hat{F}^{t,t_0}_\lambda \}_\lambda$ is the set of linearly independent operators. 
This linear independence and the group multiplication rule of
$\hat{U}_t (\Lambda,a)$ given in \eqref{eq:rule} lead to the multiplication rule of 
$\mathcal{U}_{\lambda \lambda'} (\Lambda,a)$ as
\begin{equation} 
 \sum_{\lambda'} 
 \mathcal{U}_{\lambda \lambda'} (\Lambda',a') 
 \mathcal{U}_{\lambda' \lambda''} (\Lambda,a)
 =\mathcal{U}_{\lambda \lambda''} (\Lambda'\Lambda,\Lambda a +a'). 
\label{eq:rule2}
\end{equation}
Hence, the unitary matrix 
$\mathcal{U}(\Lambda,a)$ with the components 
$\mathcal{U}_{\lambda \lambda'}(\Lambda,a)$ is a representation of the Poincar\'{e} group. 
Eq. \eqref{eq:U_t=0} helps us to simplify the invariance condition Eq.\eqref{eq:P-cov2} on the Kraus operators.
Defining the Kraus operators $\hat{E}^{t,t_0}_{\lambda}$ as 
\begin{equation}
\hat{E}^{t,t_0}_{\lambda}=e^{i\hat{H}t} \hat{F}^{t,t_0}_{\lambda} e^{-i\hat{H}t_0}
\label{eq:E}
\end{equation}
which have the completeness condition, 
\begin{equation}
\sum_{\lambda} \hat{E}^{t,t_0 \dagger}_{\lambda} \hat{E}^{t,t_0}_{\lambda}=\hat{\mathbb{I}}
\label{eq:EdgE},
\end{equation}
we can rewrite Eq.\eqref{eq:P-cov2} as 
\begin{equation}
\hat{U}^\dagger_0 (\Lambda, a) \hat{\bm{E}}\hat{U}_0 (\Lambda, a)=  \mathcal{U} (\Lambda,a) \hat{\bm{E}}.
\label{eq:P-cov3}
\end{equation}
Here, we introduced the vector 
$\hat{\bm{E}}$ with the 
$\lambda$ component 
$\hat{E}^{t,t_0}_\lambda$. 
Let the dynamical map 
$\mathcal{E}_{t,t_0}$ be given by
\begin{equation}
\mathcal{E}_{t,t_0} [\rho]=\sum_\lambda \hat{E}^{t,t_0}_\lambda \rho \hat{E}^{t,t_0 \dagger}_\lambda 
\label{eq:Ets}.
\end{equation}
The condition 
\eqref{eq:P-cov3} implies that the map 
$\mathcal{E}_{t,t_0}$ is invariant under the Poincar\'{e} group in the sense that 
\begin{equation}
\hat{U}_0 (\Lambda,a) \mathcal{E}_{t,t_0}[ \rho] \hat{U}^\dagger_0 (\Lambda,a)=\mathcal{E}_{t,t_0}[\hat{U}_0 (\Lambda,a) \, \rho \, \hat{U}^\dagger_0 (\Lambda,a)].
\label{eq:covE}
\end{equation}
Then, the dynamical map 
$\Phi_{t,t_0}$ is written with the unitary map 
$\mathcal{U}_{t,t_0}$ and the dynamical map
$\mathcal{E}_{t,t_0}$ as 
\begin{align}
\Phi_{t,t_0}[\rho]
&
=\sum_{\lambda} \hat{F}^{t,t_0}_\lambda \rho \hat{F}^{t,t_0 \dagger}_\lambda
\nonumber 
\\
&
=e^{-i\hat{H}t} \sum_{\lambda} \hat{E}^{t,t_0}_\lambda e^{i\hat{H}t_0} \rho e^{-i\hat{H}t_0}\hat{E}^{t,t_0 \dagger}_\lambda e^{i\hat{H}t}
\nonumber 
\\
&
=e^{-i\hat{H}t} 
\mathcal{E}_{t,t_0}[ e^{i\hat{H}t_0} \rho e^{-i\hat{H}t_0}] e^{i\hat{H}t}
\nonumber 
\\
&
=e^{-i\hat{H}(t-t_0)} 
\mathcal{E}_{t,t_0}[  \rho ] e^{i\hat{H}(t-t_0)}
\nonumber 
\\
&
=\mathcal{U}_{t,t_0} \circ
\mathcal{E}_{t,t_0}[  \rho ], 
\label{eq:Phi=
UE}
\end{align}
where in the fourth equality we used the condition \eqref{eq:covE} and the fact that $e^{i\hat{H}t_0}$ is the unitary transformation representing the time translation.
Our task is to determine 
$\hat{\bm{E}}$ satisfying Eq.\eqref{eq:P-cov3} (or $\mathcal{E}_{t,t_0}$ satisfying Eq.\eqref{eq:covE}). 
The irreducible unitary representations of the Poincar\'{e} group is useful for our analysis because Eq. \eqref{eq:P-cov3} is decomposed into equations for each irreducible representation subspace.  

Let us present how to classify the unitary representations of the Poincar\'{e} group \cite{Weinberg1995}.
An arbitrary four-momentum 
$q^\mu$ is represented by a standard momentum 
$\ell^\mu$ and the Lorentz transformation matrix
$(S_q)^\mu{}_\nu $
with 
\begin{equation}
q^\mu=(S_q)^\mu{}_\nu  \ell^\nu.
\label{eq:k}
\end{equation}
The unitary matrix 
$\mathcal{U}(\Lambda,a)$ is written as
\begin{equation}
\mathcal{U}(\Lambda,a)=\mathcal{U}(I,a) \mathcal{U}(\Lambda,0)=\mathcal{T}(a)\mathcal{V}(\Lambda),
\label{eq:TV}
\end{equation}
where
$I$ is the identity matrix,
$\mathcal{U}(I,a)=\mathcal{T}(a)=e^{-iP_\mu a^\mu}$ and 
$\mathcal{U}(\Lambda,0)=\mathcal{V}(\Lambda)$. 
We define the vector 
$\bm{v}_{q,\xi}$ as
\begin{equation}
\bm{v}_{q,\xi}=N_q \mathcal{V}(S_q)\bm{v}_{\ell,\xi},
\label{eq:v}
\end{equation}
where 
$P_\mu \bm{v}_{\ell,\xi}=\ell_\mu \bm{v}_{\ell,\xi}$, 
$N_q$ is the normalization, and 
the label 
$\xi$ describes the degrees of freedom other than 
those determined by $\ell^\mu$.
The vector 
$\bm{v}_{q,\xi}$ follows the transformation rules:
\begin{align}
\mathcal{T}(a)\bm{v}_{q,\xi}
&=N_q e^{-iP^\mu a_\mu} 
\mathcal{V}(S_q)\bm{v}_{\ell,\xi}
\nonumber 
\\
&=N_q \mathcal{V}(S_q)e^{-i(S_q)^\mu{}_\nu P^\nu a_\mu} \bm{v}_{\ell,\xi}
\nonumber 
\\
&=N_q \mathcal{V}(S_q)e^{-i(S_q)^\mu{}_\nu \ell^\nu a_\mu} \bm{v}_{\ell,\xi}
\nonumber 
\\
&=N_q \mathcal{V}(S_q)e^{-iq^\mu a_\mu} \bm{v}_{\ell,\xi}
\nonumber 
\\
&=e^{-iq^\mu a_\mu}\bm{v}_{q,\xi}
\label{eq:Tv}
\end{align}
and 
\begin{align}
\mathcal{V}(\Lambda)\bm{v}_{q,\xi}
&=N_q \mathcal{V}(\Lambda)
\mathcal{V}(S_q)\bm{v}_{\ell,\xi}
\nonumber 
\\
&=N_q \mathcal{V}(\Lambda S_q) \bm{v}_{\ell,\xi}
\nonumber 
\\
&=N_q \mathcal{V}(S_{\Lambda q}) \mathcal{V}(S^{-1}_{\Lambda q} \Lambda S_q)\bm{v}_{\ell,\xi}
\nonumber 
\\
&=N_q \mathcal{V}(S_{\Lambda q})  \sum_{\xi'} \mathcal{D}_{\xi'\xi} (Q(\Lambda,q)) \bm{v}_{\ell,\xi'}
\nonumber 
\\
&=\frac{N_q}{N_{\Lambda q}}  \sum_{\xi'} \mathcal{D}_{\xi'\xi} (Q(\Lambda,q)) \bm{v}_{\Lambda q,\xi'},
\label{eq:Vv}
\end{align}
where 
$Q(\Lambda,q)=S^{-1}_{\Lambda q} \Lambda S_q$. 
The matrix 
$Q(\Lambda,q)$
satisfies 
$Q^\mu{}_\nu \ell^\nu=\ell^\mu$ and the set of such matrices forms a group called the little group. 
In Eq.\eqref{eq:Vv}, 
$\mathcal{D}_{\xi\xi'} (Q)$ forms a unitary matrix 
$\mathcal{D} (Q)$ and gives a unitary representation of the little group.
The irreducible unitary representations of the Poincar\'{e} group are classified by the standard momentum
$\ell^\mu$ and the irreducible unitary representations of the little group. 
In Table \ref{table:class}, the standard momentum
$\ell^\mu$ and the little group are listed. 
For simplicity, 
$\xi$ is regarded as the label of basis vectors of the irreducible representation subspaces of the little group.  
\begin{table}[H]
  \centering
  \begin{tabular}{|c|c|c|}
  \hline
   Standard momentum $\ell^\mu$ \quad & Little group \\
  \hline
  $\ell^\mu=[M,0,0,0], \, M>0$ \, & SO(3)  \\
  $\ell^\mu=[-M,0,0,0], \, M>0$ \,  & SO(3)   \\
  $\ell^\mu=[\kappa,0,0,\kappa], \, \kappa>0$  & ISO(2)  \\
  $\ell^\mu=[-\kappa,0,0,\kappa], \, \kappa>0$  & ISO(2)   \\
  $\ell^\mu=[0,0,0,N],\, N^2>0 $  & SO(2,1)   \\
  $\ell^\mu=[0,0,0,0] $  & SO(3,1)   \\
  \hline
  \end{tabular}
  \caption{Classification of the standard momentum $\ell^\mu$ and the little group associated with $\ell^\mu$.}
    \label{table:class}
\end{table}
We investigate Eq.\eqref{eq:P-cov3} restricted on each irreducible representation. 
For convenience, we separately focus on the Lorentz transformation and the spacetime translation in Eq.\eqref{eq:P-cov3}.  
The unitary operator 
$\hat{U}_0 (\Lambda,a)$ is written as
\begin{equation}
\hat{U}_0 (\Lambda,a)=\hat{U}_0 (I,a) \hat{U}_0 (\Lambda,0)=\hat{T}(a)\hat{V}(\Lambda),
\label{eq:hatTV}
\end{equation}
where
$\hat{U}_0 (I,a)=\hat{T}(a)=e^{-i\hat{P}_\mu a^\mu}$ with the four-momentum operator
$\hat{P}^\mu$ and 
$\hat{U}_0 (\Lambda,0)=\hat{V}(\Lambda)$ with the generators 
$\hat{\bm{J}}$ and 
$\hat{\bm{K}} (0)$. 
From Eq.\eqref{eq:P-cov3} for 
$\Lambda=I$, we have 
\begin{equation}
\hat{T}^\dagger (a) \, \hat{\bm{E}} \, \hat{T} (a) =\mathcal{T}(a) \hat{\bm{E}}. 
\label{eq:TMT}
\end{equation}
Eq.\eqref{eq:P-cov3} for  
$a^\mu=0$ gives
\begin{equation}
\hat{V}^\dagger (\Lambda) \, \hat{\bm{E}} \, \hat{V} (\Lambda) =\mathcal{V}(\Lambda) \hat{\bm{E}}.
\label{eq:VMV}
\end{equation}
Introducing 
$\hat{E}_{q,\xi}=\bm{v}^\dagger_{q,\xi} \hat{\bm{E}}$, we obtain the following equations from Eqs.\eqref{eq:TMT} and \eqref{eq:VMV}:
\begin{equation}
\hat{T}^\dagger (a) \hat{E}_{q,\xi} \hat{T} (a) =e^{-iq_\mu a^\mu} \hat{E}_{q,\xi}
\label{eq:TMT2}
\end{equation}
and 
\begin{equation}
\hat{V}^\dagger (\Lambda) \hat{E}_{q,\xi} \hat{V} (\Lambda) =\frac{N^*_q}{N^*_{\Lambda^{-1} q} } \sum_{\xi'} \mathcal{D}^*_{\xi'\xi}(Q(\Lambda^{-1},q))  \hat{E}_{\Lambda^{-1}q,\xi'},
\label{eq:VMV2}
\end{equation}
where we used Eqs.\eqref{eq:Tv} and \eqref{eq:Vv}, and 
$Q(\Lambda, q)=S^{-1}_{\Lambda q} \Lambda S_q$. 
The label 
$\xi$ can take discrete or continuous values. 
For the continous case, the summation 
$\sum_\xi$ is replaced with 
the integration 
$\int d\mu(\xi)$ with a measure 
$\mu(\xi)$. 
Focusing on Eq.\eqref{eq:VMV2} for 
$\Lambda=S_q$,  we get
\begin{equation}
\hat{V}^\dagger (S_q) \hat{E}_{q,\xi} \hat{V} (S_q) =N^*_q \, \hat{E}_{\ell,\xi},
\label{eq:SMS}
\end{equation}
where note that
$N_\ell=1$ and 
$Q(S^{-1}_q,q)=S^{-1}_{S^{-1}_q q} S^{-1}_q S_q=S^{-1}_{\ell}=I$ hold by the definition of
$\bm{v}_{q,\xi}$. 
Eq.\eqref{eq:SMS} tells us that the Kraus operators 
$\hat{E}_{q,\xi}$ is determined from the Kraus operators $\hat{E}_{\ell,\xi}$ with the standard momentum 
$\ell^\mu$. 
All we have to do is to give the form of the Kraus operators
$\hat{E}_{\ell,\xi}$. 
To this end, we present the following equations given by Eq.\eqref{eq:TMT2} for 
$q^\mu=\ell^\mu$ and by Eq.\eqref{eq:VMV2} for 
$q^\mu=\ell^\mu$ and 
$\Lambda=W$ with 
$W^\mu{}_\nu \ell^\nu=\ell^\mu$, respectively:
\begin{align}
\hat{T}^\dagger (a) \hat{E}_{\ell,\xi} \hat{T} (a) &=e^{-i\ell_\mu a^\mu} \hat{E}_{\ell,\xi}
\label{eq:TMT3},
\\
\hat{V}^\dagger (W) \hat{E}_{\ell,\xi} \hat{V} (W) &= \sum_{\xi'} \mathcal{D}^*_{\xi'\xi}(W^{-1})  \hat{E}_{\ell,\xi'},
\label{eq:WMW}
\end{align}
where 
$Q(\Lambda^{-1},q)=Q(W^{-1},\ell)=S^{-1}_{W^{-1} \ell} W^{-1} S_\ell =W^{-1}$.
In the next section, we construct a model of the dynamical map with the Poincar\'{e} invariance to describe the reduced dynamics of a spinless massive particle.

\section{A model of the dynamical map for a spinless massive particle}
\label{sec:4} 

In this section, based on Eqs.\eqref{eq:TMT3} and \eqref{eq:WMW}, we give a model of the dynamical map with the Poincar\'{e} invariance. 
To simplify the analysis, we consider a spinless particle with a mass 
$m$ and its  Hilbert space 
$\mathcal{H}_0 \oplus \mathcal{H}_1$, where 
$\mathcal{H}_0$ is the one-dimensional Hilbert space with the vacuum state 
$|0\rangle$ and 
$\mathcal{H}_1$ is the irreducible subspace with one-particle states. 
A state vector 
$|\Psi \rangle$ in 
$\mathcal{H}_1$ 
($\, |\Psi \rangle \in \mathcal{H}_1 \,$) is 
\begin{equation}
|\Psi \rangle = \int d^3p  \Psi(\bm{p}) \, \hat{a}^\dagger (\bm{p})|0 \rangle,
\label{eq:Psi}
\end{equation}
where the vacuum state
$|0\rangle$ satisfies
$\hat{a}(\bm{p})|0\rangle=0$, 
$\Psi(\bm{p})$ with the momentum 
$\bm{p}$ is the wave function, 
$\hat{a}(\bm{p})$ and 
$\hat{a}^\dagger (\bm{p}) $ are the annihilation and creation operators with
\begin{equation}
[\hat{a}(\bm{p}), \hat{a}(\bm{p}')]=0=[\hat{a}^\dagger(\bm{p}), \hat{a}^\dagger(\bm{p}')] 
\quad 
[\hat{a}(\bm{p}), \hat{a}^\dagger(\bm{p}')]=\delta^3(\bm{p}-\bm{p}').
\label{eq:aadg}
\end{equation}
Here, 
$[\hat{A},\hat{B}]$ is the commutator,
$[\hat{A},\hat{B}]=\hat{A}\hat{B} - \hat{B} \hat{A}$.
In Ref.\cite{Weinberg1995, Peres2004}, the transformation rules of 
$\hat{a}^\dagger (\bm{p})$ are given by
\begin{align}
\hat{T}(a) \hat{a}^\dagger (\bm{p}) \hat{T}^\dagger (a)
&=e^{-ip^\mu a_\mu} \hat{a}^\dagger (\bm{p}), 
\label{eq:TaT}
\\
\hat{V}(\Lambda) \hat{a}^\dagger (\bm{p}) \hat{V}^\dagger (\Lambda)
&=
\sqrt{\frac{E_{\bm{p}_\Lambda}}{E_{\bm{p}}}}
 \hat{a}^\dagger (\bm{p}_\Lambda), 
\label{eq:VaV}
\end{align}
where 
$E_{\bm{p}}=p^0=\sqrt{\bm{p}^2+m^2}$, 
$E_{\bm{p}_\Lambda}=(\Lambda p)^0$ and 
$\bm{p}_\Lambda$ is the vector with the components 
$(\bm{p}_\Lambda)^i=(\Lambda p)^i$. 

We consider the Kraus operators
$\hat{E}_{\ell,\xi}$ acting on the Hilbert space $\mathcal{H}_0 \oplus \mathcal{H}_1$, that is, $\hat{E}_{\ell,\xi}: \mathcal{H}_0 \oplus \mathcal{H}_1 \rightarrow \mathcal{H}_0 \oplus \mathcal{H}_1 $, which have the following form 
\begin{equation}
\hat{E}_{\ell,\xi}=A_{\ell,\xi}\hat{\mathbb{I}}+\int d^3 p  \, B_{\ell,\xi} (\bm{p}) \hat{a}(\bm{p})+\int d^3 p' d^3p \, C_{\ell,\xi} (\bm{p}', \bm{p}) \hat{a}^\dagger (\bm{p}') \hat{a}(\bm{p}).
\label{eq:ABC}
\end{equation}
The dynamical map given by these operators describes the reduced dynamics of the particle, which can decay into the vacuum state. 
Substituting Eq.\eqref{eq:ABC} into  Eq.\eqref{eq:TMT3} and 
Eq.\eqref{eq:WMW}, we obtain 
\begin{align}
&A_{\ell,\xi}=e^{-i\ell^\mu a_\mu} A_{\ell,\xi}, 
\label{eq:TAT}
\\
&B_{\ell,\xi} (\bm{p}) e^{-ip^\mu a_\mu} =B_{\ell,\xi} (\bm{p}) e^{-i\ell^\mu a_\mu},
\label{eq:TBT}
\\
&C_{\ell,\xi} (\bm{p}', \bm{p}) e^{i({p'}^\mu-p^\mu) a_\mu}=C_{\ell,\xi} (\bm{p}', \bm{p}) e^{-i\ell^\mu a_\mu}
\label{eq:TCT},
\end{align}
and 
\begin{align}
&A_{\ell,\xi}=\sum_{\xi'} \mathcal{D}^*_{\xi'\xi}(W^{-1})  A_{\ell,\xi'}, 
\label{eq:WAW}
\\
&\sqrt{\frac{E_{\bm{p}_W}}{E_{\bm{p}}}}\sum_{\sigma} B_{\ell,\xi} (\bm{p}_W)    =\sum_{\xi'}\mathcal{D}^*_{\xi'\xi}(W^{-1}) B_{\ell,\xi'} (\bm{p}),
\label{eq:WBW}
\\
&\sqrt{\frac{E_{\bm{p}'_W}E_{\bm{p}_W}}{E_{\bm{p}'}E_{\bm{p}}}}
 C_{\ell,\xi} (\bm{p}'_W, \bm{p}_W)
=
\sum_{\xi'}\mathcal{D}^*_{\xi'\xi}(W^{-1})
C_{\ell,\xi'} (\bm{p}', \bm{p}) 
\label{eq:WCW}.
\end{align}
The derivation of these equations is devoted in Appendix \ref{App:ABC}.

We can analyze the form of
$A_{\ell,\xi}$,
$B_{\ell,\xi} (\bm{p})$ 
and 
$C_{\ell,\xi} (\bm{p}',\bm{p})$
for a spinless massive particle. 
The long computations presented in Appendices \ref{App:massive} give the dynamical map with the Poincar\'{e} invariance, 
$\Phi_{t,t_0}[\rho(t_0)]=\mathcal{U}_{t,t_0} \circ \mathcal{E}_{t,t_0}[\rho(t_0)]$, with the unitary map
$\mathcal{U}_{t,t_0}$ given in \eqref{eq:uts} and $\mathcal{E}_{t,t_0}$ as
\begin{align}
\mathcal{E}_{t,t_0}[\rho(t_0)] 
&=\beta_{t,t_0} \int d^3 p  \, \hat{a}(\bm{p}) \rho(t_0) \hat{a}^\dagger (\bm{p})+  \Big(
 \hat{\mathbb{I}}+\gamma_{t,t_0} \hat{N}
\Big) 
\rho(t_0) \Big(
 \hat{\mathbb{I}}+ \gamma_{t,t_0} \hat{N}
\Big)^\dagger
\nonumber 
\\
&
+\int d^3p \int d^3q \, \delta_{t,t_0} (p,q)
\, \hat{a}^\dagger(\bm{p})\hat{a}(\bm{p}) 
\, \rho(t_0) \, 
\hat{a}^\dagger (\bm{q})\hat{a}(\bm{q}),
\label{eq:map}
\end{align}
where 
$\hat{N}$ is the number operator defined by 
\begin{equation}
\hat{N}=\int d^3p \,  \hat{a}^\dagger(\bm{p}) \hat{a}(\bm{p}). 
\label{eq:N}
\end{equation}
The function 
$\delta_{t,t_0} (p,q)$
is non-negative and Lorentz invariant in the sense that
\begin{equation}
\int d^3p d^3q f^*(p) \delta_{t,t_0} (p,q) f(q) \geq 0, \quad \delta_{t,t_0}(\Lambda p, \Lambda q) = \delta_{t,t_0}(p,q),
\label{eq:delta}
\end{equation}
and hence 
$\delta_{t,t_0} (p,p)$ does not depend on the three-momentum of the particle. 
In the following, 
$\delta_{t,t_0}(p,p)$ is simply denoted by
$\delta_{t,t_0}$. 
The parameters
$\beta_{t,t_0} $, 
$\gamma_{t,t_0}$ and
$\delta_{t,t_0} $
satisfy 
\begin{equation}
\beta_{t,t_0} \geq 0, \quad \beta_{t,t_0}+\gamma^*_{t,t_0}+\gamma_{t,t_0}+|\gamma_{t,t_0}|^2 +\delta_{t,t_0}=0,
\label{eq:comp0}
\end{equation}
where 
the former inequality is required from the fact that the density operator 
$\mathcal{E}_{t,t_0}[\rho(t_0)]$ should be positive, and the latter condition in \eqref{eq:comp0} is yielded from the completeness of the Kraus operators \eqref{eq:FdgF}. 

From the transformation rules of the creation and the annihilation operators, Eqs.\eqref{eq:TaT} and 
\eqref{eq:VaV}, we can check that the map 
$\mathcal{E}_{t,t_0}$ satisfies the invariance condition \eqref{eq:covE}.
Since the unitary map 
$\mathcal{U}_{t,t_0}$ is invariant under the Poincar\'{e} group, which is discussed around Eq.\eqref{eq:uts}, 
we can confirm that 
$\Phi_{t,t_0}$ is also invariant.

We discuss the conservation laws of the Poincar\'{e} generators, 
$\hat{H}$, $\hat{\bm{P}}$, $\hat{\bm{J}}$, and 
$\hat{\bm{K}}(t)$ 
in quantum mechanics, which means that the all-order moments of each operator are conserved during time evolution. 
For this purpose, it is useful to consider the characteristic function,
\begin{equation}
\chi_t (\theta,a) =\text{Tr}[e^{-ia_\mu \hat{P}^\mu +\frac{i}{2}\theta_{\mu \nu} \hat{J}^{\mu\nu}(t)} \rho(t)],
\label{eq:chit}
\end{equation}
where 
$\hat{P}^\mu=[\hat{H},\hat{\bm{P}}]$ is the four-momentum, and 
the antisymmetric tensor 
$\hat{J}^{\mu\nu}(t)$ is given by
$\hat{J}^{jk}(t)=\epsilon^{ijk}\hat{J}_i$ and 
$J^{i0}(t)=\hat{K}^i(t)$, which depends on time in the Schr\"{o}dinger picture. 
For example, the 
$n$-th order moment of the energy is given as
$\text{Tr}[\hat{H}^n \rho(t)]=(-i)^n \partial^n_{a^0} \chi_t (\theta,a)|_{\theta=0=a}$.
The time evolution of the density operator is 
$\rho(t)=\Phi_{t,t_0}[\rho(t_0)]=\mathcal{U}_{t,t_0} \circ \mathcal{E}_{t,t_0}[\rho(t_0)]$. 

Let us investigate the conservation of the four-momentum 
$\hat{P}^\mu$ for the above model of the massive particle. 
The characteristic function 
$\chi_t (0,a)$ is computed as 
\begin{align}
\chi_t (0,a)
&=\text{Tr}[e^{-ia_\mu \hat{P}^\mu } \rho(t)]
\nonumber 
\\
&
=\chi_{t_0} (0,a)
+\beta_{t,t_0}\text{Tr}\Big[ \hat{N}\Big(\hat{\mathbb{I}}-e^{-ia_\mu \hat{P}^\mu} \Big)\rho(t_0) \Big].
\label{eq:chit2}
\end{align}
We thus find that the energy of the particle is not conserved, $\chi_t (0,a)\neq\chi_{t_0} (0,a)$, even when the map is invariant under the Poincar\'{e} group. 
Such a deviation between symmetry and conservation law was discussed in, for example, Refs \cite{Cirstoiu2020} and \cite{Marvian2014}. 
If the parameter 
$\beta_{t,t_0}$ vanishes, then 
$\chi_t (0,a)=\chi_s(0,a)$ and hence the energy is conserved.
For such a case, we find the dynamical map
$\Phi_{t,t_0}$ with the Poincar\'{e} invariance, which guarantees the four-momentum conservation.

Under the four-momentum conservation (the condition 
$\beta_{t,t_0}=0$), we further examine the conservation of 
$\hat{J}^{\mu\nu}(t)$. 
The characteristic function 
$\chi_t (\theta,0)$ is 
\begin{align}
\chi_t (\theta,0)
&=\text{Tr}[e^{\frac{i}{2} \theta_{\mu \nu} \hat{J}^{\mu \nu}(t) } \rho(t)]
\nonumber 
\\
&
=\chi_{t_0} (\theta,0)-\delta_{t,t_0}\text{Tr}\Big[ \hat{N} e^{\frac{i}{2} \theta_{\mu \nu} \hat{J}^{\mu \nu}(t_0)} \rho(t_0) \Big]+ \text{Tr}\Big[
\int d^3p\, \delta_{t,t_0}(p,\Lambda p) \hat{a}^\dagger(\bm{p})\hat{a}(\bm{p}) 
\, e^{\frac{i}{2} \theta_{\mu \nu} \hat{J}^{\mu \nu}(t_0)}\rho(t_0) \Big],
\label{eq:chit3}
\end{align}
where 
$\Lambda=\Lambda(\theta)$ is the Lorentz transformation matrix determined by 
$\theta_{\mu\nu}$. 
To conserve 
$\hat{J}^{\mu\nu}(t)$, 
$\delta_{t,t_0}=\delta_{t,t_0}(p,\Lambda p)$ should hold for all 
$\Lambda$, and hence 
$\delta_{t,t_0}(p,q)=\delta_{t,t_0}$. 
The dynamical map 
$\mathcal{E}_{t,t_0}$ for a spinless massive particle becomes
\begin{equation}
\mathcal{E}_{t,t_0}[\rho(t_0)] 
= \Big(
 \hat{\mathbb{I}}+\gamma_{t,t_0} \hat{N} 
\Big) 
\rho(t_0) \Big(
 \hat{\mathbb{I}}+\gamma_{t,t_0} \hat{N}
\Big)^\dagger
+\delta_{t,t_0}  \hat{N} \rho(t_0) \hat{N}.
\label{eq:spinless2}
\end{equation}
When the density
operator 
$\rho(t_0)$ is given by one-particle states, 
we have
$\hat{N}\rho(t_0)=\rho(t_0)=\rho(t_0) \hat{N}$, and then the dynamical map 
$\Phi_{t,t_0}$ with the Poincar\'{e} invariance is
\begin{equation}
\Phi_{t,t_0}[\rho(t_0)] 
= \mathcal{U}_{t,t_0}\circ \mathcal{E}_{t,t_0}[\rho(t_0)]=(|1+\gamma_{t,t_0}|^2+\delta_{t,t_0})\mathcal{U}_{t,t_0}[\rho(t_0)]=\mathcal{U}_{t,t_0}[\rho(t_0)],
\label{eq:spinless3}
\end{equation}
where we used the condition 
$\delta_{t,t_0}=-\gamma_{t,t_0}-\gamma^*_{t,t_0 }-|\gamma_{t,t_0}|^2$ given by setting 
$\beta_{t,t_0}=0$ for the second equation in \eqref{eq:comp0}.
Hence, the
dynamical map with the Poincar\'{e} invariance for a spinless massive particle is reduced to the unitary map when the conservation of the Poincar\'{e} generators holds. 
The result corresponds to an extension of the analysis in \cite{TorosPhD}.

\section{A covariant formulation}
\label{sec:extra} 

In the previous section, we get the form of the dynamnical map with the Poincar\'{e} invariance on a Minkowski time in a special coordinate system. 
In the following, let us rewrite the previous formulation in a covariant way. 
We consider the foliation of Cauchy surfaces 
$\{\Sigma_\tau \}_\tau$ and the unit timelike vector 
$n^\mu$ normal to 
$\Sigma_\tau$, where the parameter 
$\tau$ is generated by 
$n^\mu$.  
For example, the parameter 
$\tau$ is defined by 
$dx^\mu=n^\mu d\tau$ with a constant 
$n^\mu$ in the Minkowski space.
Letting 
$x^\mu_0$ be a constant of integration, 
the solution of the equation yields
$\tau=-n_\mu(x^\mu -x^\mu_0)$, which is invariant in any inertial coordinate system.
In the present formulation, the density operator 
$\rho(\tau)$ of a quantum system is defined on the Cauchy surface 
$\Sigma_{\tau}$. 
The change of the density operator from a Cauchy surface 
$\Sigma_{\tau_0}$ to another 
$\Sigma_{\tau}$ is regarded as the time evolution of the density operator from 
$\tau_0$ to 
$\tau \, (>\tau_0)$.

Since the parameter 
$\tau$ is coordinate invariant, the Cauchy surface 
$\Sigma_\tau$ given as a 
$\tau=\text{constant}$ hypersurface does not depend on any choice of an inertial coordinate system.
On the other hand, the density operator depends on it. 
This is because the density operator of a quantum system is specified by the statistical outcome of observables such as a momentum and a spin.  
When the density operator in one inertial coordinate system is 
$\rho(\tau)$ on 
$\Sigma_\tau$, the density operator of the same quantum system in another inertial coordinate system is given as
$\hat{U}_\tau \rho(\tau) \hat{U}^\dagger_\tau$ on the same Cauchy surface 
$\Sigma_\tau$. 
The two coordinate systems are connected by a Poincar\'{e} transformation, which is represented as the unitary operator 
$\hat{U}_\tau$ acting on the density operator.
In the Schr\"{o}dinger picture, the Poincar\'{e} generators of
$\hat{U}_\tau$ are (see also Ref.\cite{Fleming1966})
\begin{equation}
\hat{\Theta}=-n_\mu \hat{P}^\mu, 
\quad \hat{\Pi}^\mu=\hat{P}^\mu-n^\mu \hat{\Theta}, 
\quad \hat{L}^\mu =\frac{1}{2} \epsilon^{\mu\alpha \beta\gamma} \hat{J}_{\alpha \beta} n_\gamma
\quad \hat{N}^\mu=\hat{J}^{\mu \nu}n_\nu,
\label{eq:ThPiLN}
\end{equation}
which satisfy the commutation relations,
\begin{align}
[\hat{\Pi}_\mu, \hat{\Pi}_\nu] &=0, 
\label{eq:PiPi}
\\
[\hat{\Pi}_\mu, \hat{\Theta}] &=0,
\label{eq:PiTh}
\\
[\hat{L}_\mu, \hat{\Theta}] &=0,
\label{eq:LTh}
\\
[\hat{L}_\mu, \hat{L}_\nu]&=i\epsilon_{\mu\nu\alpha\beta} n^\alpha \hat{L}^\beta ,
\label{eq:LL}
\\
[\hat{L}_\mu, \hat{\Pi}_\nu]&=i\epsilon_{\mu\nu\alpha\beta} n^\alpha \hat{\Pi}^\beta,
\label{eq:LPi}
\\
[\hat{L}_\mu, \hat{N}_\nu]&=i\epsilon_{\mu\nu\alpha\beta} n^\alpha \hat{N}^\beta,
\label{eq:LN}
\\
[\hat{N}_\mu, \hat{\Pi}_\nu]&=i (\eta_{\mu\nu}+n_\mu n_\nu)\hat{\Theta},
\label{eq:NPi}
\\
[\hat{N}_\mu, \hat{\Theta}]&=i \hat{\Pi}_\mu,
\label{eq:NTh}
\\
[\hat{N}_\mu, \hat{N}_\nu]&=-i\epsilon_{\mu\nu\alpha\beta} n^\alpha \hat{L}^\beta.
\label{eq:NN}
\end{align}

In the foliation with 
$n^\mu=[1,0,0,0]$ in a coordinate, the generators 
$\hat{\Theta}$, 
$\hat{\Pi}^\mu$, 
$\hat{L}^\mu$ and
$\hat{N}^\mu$ are with 
$\hat{\Theta}=\hat{H}$, 
$\hat{\Pi}^\mu=[0,\hat{\bm{P}}]$,  
$\hat{L}^\mu=[0,\hat{\bm{J}}]$ and
$\hat{N}^\mu=[0, \hat{\bm{K}}]$, respectively. 
These are nothing but the Poincar\'{e} generators for the Minkowski time considered until the previous section. 
Since each generator defined with the Cauchy surface 
$\Sigma_\tau$ is conserved in the flow by the system Hamiltonian
$\hat{\Theta}$, the equation  $d\hat{N}^\mu_\text{H}/d\tau=0$ with 
$\hat{N}^\mu_\text{H}(\tau)=e^{i\hat{\Theta}\tau}\hat{N}^\mu (\tau)e^{-i\hat{\Theta}\tau}$ holds. 
This suggests that the operator 
$\hat{N}^\mu$ depends on 
$\tau$ as
\begin{equation}
\hat{N}^\mu (\tau)=e^{-i\hat{\Theta}\tau}\hat{N}^\mu (0)e^{i\hat{\Theta}\tau}
\label{eq:N(0)}
\end{equation}
in the Schr\"{o}dinger picture.
The similar discussion on the time dependence of 
$\hat{\bm{K}}(t)$ has been devoted around Eq.\eqref{eq:KK_0} in Sec.\ref{sec:3}. 
Then, the 
$\tau$ dependence of the unitary operator 
$\hat{U}_\tau (\Lambda,a)$ generated by 
$\hat{\Theta}$, 
$\hat{\Pi}^\mu$, 
$\hat{L}^\mu$ and
$\hat{N}^\mu (\tau)$ is given by
\begin{equation}
\hat{U}_\tau (\Lambda,a)=e^{-i\hat{\Theta}\tau} \hat{U}_0 (\Lambda,a)e^{i\hat{\Theta}\tau},
\label{eq:U(0)}
\end{equation}
where the commutation relations \eqref{eq:PiTh} and 
\eqref{eq:LTh} were used, and 
the unitary operator 
$\hat{U}_0 (\Lambda,a)$ is generated by 
$\hat{\Theta}$, 
$\hat{\Pi}^\mu$, 
$\hat{L}^\mu$ and
$\hat{N}^\mu(0)$.

To manifest covariance, we use the Lorentz invariant measure 
$d\mu(p)$ with 
$d\mu(p)=d^4p \, \delta(p^2+m^2)\theta(p^0)$ for a massive particle.
Also, we introduce the annihilation and creation operators 
$\hat{A}(p)$ and 
$\hat{A}^\dagger(p)$ satisfying 
\begin{equation}
[\hat{A}(f), \hat{A}(g)]=0=[\hat{A}^\dagger(f), \hat{A}^\dagger(g)], 
\quad 
[\hat{A}(f), \hat{A}^\dagger(g)]=\int d\mu(p) \,f(p) g^*(p),
\label{eq:AAdg}
\end{equation}
where 
$f(p)$ and 
$g(p)$ are a complex function, and 
\begin{equation}
\hat{A}(f)=\int d\mu(p) \, f(p) \hat{A}(p).
\label{eq:A(f)}
\end{equation}
The Poincar\'{e} transformation rule of 
$\hat{A}^\dagger(p)$ is 
\begin{align}
\hat{U}_0(\Lambda,a) \hat{A}^\dagger (p) \hat{U}^\dagger_0(\Lambda,a)
&=e^{-i(\Lambda p)^\mu a_\mu}  \hat{A}^\dagger (\Lambda p).
\label{eq:UAU}
\end{align}
In the previous formulation, the annihilation operator 
$\hat{A}(p)$ corresponds to
$\sqrt{E_{\bm{p}}}\hat{a}(\bm{p})$.
We can use the same procedure in the previous sections and Appendix \ref{App:massive} to write down the dynamical map with the Poincar\'{e} invariance in a covariant way. 
The dynamical map 
$\Phi_{\tau,\tau_0}=\mathcal{U}_{\tau,\tau_0} \circ \mathcal{E}_{\tau,\tau_0}$ for a massive 
spinless particle is given by 
\begin{equation}
\mathcal{U}_{\tau,\tau_0}[\rho]=e^{-i\hat{\Theta}(\tau-\tau_0)} \, \rho \,e^{i\hat{\Theta}(\tau-\tau_0)},
\label{eq:Uincov}
\end{equation}
and 
\begin{align}
\mathcal{E}_{\tau,\tau_0}[\rho] 
&= \beta_{\tau,\tau_0} \int d\mu(p)  \hat{A}(p) \, \rho \, \hat{A}^\dagger (p)+ \Big(
 \hat{\mathbb{I}}+\gamma_{\tau,\tau_0} \hat{N} 
\Big) 
\, \rho \, \Big(
 \hat{\mathbb{I}}+ \gamma_{\tau,\tau_0} \hat{N} 
\Big)^\dagger
\nonumber 
\\
&
+\int d\mu(p) d\mu(q) \, \delta_{\tau, \tau_0}(p,q) \hat{A}^\dagger (p) \hat{A}(p) \, \rho \, \hat{A}^\dagger (q) \hat{A}^\dagger (q), 
\label{eq:Eincov}
\end{align}
where the number operator 
$\hat{N}$ is
\begin{equation}
\hat{N} =\int d\mu(p) \hat{A}(p)\hat{A}^\dagger(p).
\label{eq:Nincov}
\end{equation}
For the Minkowski time in a special coordinate, in which
$n^\mu=[1,0,0,0]$, 
$\tau=t$ and 
$\tau_0=t_0$, the above dynamical map is reduced to that derived in the previous section.

We give a comment on what may happen if the obtained dynamics are described in another foliation of Cauchy surfaces
$\{\tilde{\Sigma}_\lambda \}_{\lambda}$ with a normal vector 
$m^\mu=\Lambda^{\mu}{}_\nu n^\nu$, where 
$\Lambda^\mu{}_\nu$ is a Lorentz transforamtion matrix.
Here, note that 
$m^\mu$ is not given by a Lorentz (coordinate) transformation but is a different objective vector from 
$n^\mu$. 
In the another foliation, the  may give the action at distance, that is, may induce the violation of causality. 
To investigate this conceptual problem clearly, the local description of the  would be required. 
This is because we should carefully specify each description of the  in different foliations by a local quantity at a spacetime point in the intersection of two Cauchy surfaces 
$\Sigma_\tau$ and
$\tilde{\Sigma}_\lambda$.
Elucidating what constraints to the present framework are derived from causality and locality remains a challenge for the future.

\section{Conclusion}
\label{sec:5}

We discussed what a dynamical map describing the reduced  of an open quantum system is realized under the Poincar\'{e} symmetry. 
The unitary representation theory of the Poincar\'{e} group refines the condition for the dynamical map with the Poincar\'{e} nvariance. 
We derived the model of the dynamical map for a spinless massive particle. 
In the model, the particle can decay into the vacuum state, and we found a model of the dynamical map with the four-momentum conservation. 
Further, we showed that the map is unitary under the conservation of the Poincar\'{e} generators, if the map is restricted on density operators of one-particle states of the spinless massive particle.
In this way, it was exemplified that the Poincar\'{e} symmetry strongly constrains the possible dynamics of an open quantum system. 
We also formulated the dynamical map with the Poincar\'{e} invariance in a covariant way.
In the present analysis, we did not clarify what total dynamics of the system and the environment lead to the dynamical map obtained here. 
This is devoted as a future work.

In this paper, we assumed an open system with a single particle. 
Our analysis is possible to be extended to the case with many particles. 
Considering interactions among many particles, we can understand more general effective theories of open quantum systems in terms of the Poincar\'{e} symmetry. 
For the particles interacting via gravity, we can also discuss the models with intrinsic gravitational decoherence, which have been proposed in \cite{Kafri2014,Diosi1987,Diosi1989,Penrose1996,Tilloy2016}.
These models are written in the theory of open quantum systems. 
In the weak field regime of gravity, the Poincar\'{e} symmetry may provide a guidance for establishing the theory of gravitating particles.

This paper has the potential to develop a relativistic theory of open quantum systems. 
To describe the reduced dynamics of an open quantum system, a Markovian quantum master equation is often adopted. 
How such a master equation is consistent with relativity has been discussed \cite{Alicki1986, Diosi2022, Meng2021}. 
Applying the present approach, it will be possible to discuss the quantum Markov dynamics with the Poincar\'{e} invariance.
In doing so, it is also worth considering the description of the dynamics in quantum field theory and examining relativistic causality.
The previous works \cite{Sorkin1993,Fewster2020,Bostelmann2021,Fewster2022} discussed the consistency between the measurements of local observables and the relativistic causality in quantum field theory. 
It may be interesting to understand the dynamics of an open quantum system with the Poincar\'{e} symmetry as a causal measurement process by an environment.

\begin{acknowledgements}
We thank Y. Kuramochi for useful discussions and comments related to this paper. 
A.M. was supported by 2022 Research Start Program 202203.
\end{acknowledgements}

\begin{appendix}

\section{Derivation of Eq.\eqref{eq:opsum}}
\label{App:opsum}

We here give a derivation of Eq.\eqref{eq:opsum} starting from Eq.\eqref{eq:Phi}. 
For simplicity, we assume that the initial state of the environment is pure, 
$\rho_\text{E}=|0\rangle_\text{E} \langle 0|$. 
We then find
\begin{align}
\rho(t)
&=\text{Tr}_\text{E}[\hat{U}(t,t_0) \rho(t_0) \otimes |0\rangle_\text{E} \langle 0|\, \hat{U}^\dagger (t,t_0)] 
\nonumber 
\\
&=\sum_{n} {}_\text{E} \langle n| \hat{U}(t,t_0) \rho(t_0) \otimes |0\rangle_\text{E} \langle 0| \, \hat{U}^\dagger (t,t_0) |n \rangle_\text{E}
\nonumber 
\\
&=\sum_{n} \Big({}_\text{E} \langle n| \hat{U}(t,t_0) |0\rangle_\text{E}\Big) \rho(t_0) \Big({}_\text{E} \langle 0| \hat{U}^\dagger (t,t_0) |n \rangle_\text{E} \Big)
\nonumber 
\\
&=\sum_n \hat{F}^{t,t_0}_n \rho(t_0)\hat{F}^{t,t_0 \dagger}_n,
\label{eq:PhiApp} 
\end{align}
where 
$\hat{F}^{t,t_0}_n ={}_\text{E} \langle n| \hat{U}(t,t_0) |0\rangle_\text{E}$ is the Kraus operator acting on the system density operator 
$\rho(t_0)$. 
Replacing 
$n$ with 
$\lambda$, we get Eq.\eqref{eq:opsum}.

Even when the environment is initially in a mixed state 
$\rho_\text{E}$, for example, a thermal state, we can derive the operator-sum representation \eqref{eq:opsum}. 
Eq.\eqref{eq:Phi} is rewritten as 
\begin{align}
\rho(t)
&=\text{Tr}_\text{E}[\hat{U}(t,t_0) \rho(t_0) \otimes \rho_\text{E} \, \hat{U}^\dagger (t,t_0)] 
\nonumber 
\\
&=\text{Tr}_\text{E}[\hat{U}(t,t_0) \rho(t_0) \otimes \sqrt{\rho_\text{E}} \sqrt{\rho_\text{E}} \, \hat{U}^\dagger (t,t_0)] 
\nonumber 
\\
&=\text{Tr}_\text{E}[\hat{U}(t,t_0) \rho(t_0) \otimes \sqrt{\rho_\text{E}} \sum_m |m\rangle_\text{E} \langle m| \sqrt{\rho_\text{E}} \, \hat{U}^\dagger (t,t_0)] 
\nonumber 
\\
&=\sum_n {}_\text{E} \langle n| \hat{U}(t,t_0) \rho(t_0) \otimes \sqrt{\rho_\text{E}} \sum_m |m\rangle_\text{E} \langle m| \sqrt{\rho_\text{E}} \, \hat{U}^\dagger (t,t_0) |n \rangle_\text{E}
\nonumber 
\\
&=\sum_{n,m} \Big( {}_\text{E} \langle n| \hat{U}(t,t_0) \sqrt{\rho_\text{E}} |m\rangle_\text{E} \Big) \rho(t_0)   \Big({}_\text{E} \langle m| \sqrt{\rho_\text{E}} \, \hat{U}^\dagger (t,t_0) |n \rangle_\text{E} \Big)
\nonumber 
\\
&=\sum_{n,m} \hat{F}^{t,t_0}_{n,m} \rho(t_0)  \hat{F}^{t,t_0 \dagger}_{n,m},
\label{eq:PhiApp2} 
\end{align}
where in the second line we used the fact that 
$\rho_\text{E}$ is a non-negative operator and uniquely has its square root 
$\sqrt{\rho_\text{E}}$, and in the third line we inserted the completness relation 
$\sum_{m}|m\rangle_\text{E} \langle m|=\hat{\mathbb{I}}_\text{E}$. 
Replacing the label 
$(n,m)$ of the Kraus operator 
$\hat{F}^{t,t_0}_{n,m} = {}_\text{E} \langle n| \hat{U}(t,t_0) \sqrt{\rho_\text{E}} |m\rangle_\text{E}$ acting on 
$\rho(t_0)$ with 
$\lambda$, we get Eq.\eqref{eq:opsum} again.
Here, we assumed the initial uncorrelated state of the system and the environment. In the theory of an open quantum system, we may consider an initial state with correlation, which may lead to a more general reduced dynamics of the system (for example, see the review  \cite{Breuer2016}). 
In this paper, we do not discuss such dynamics.

\section{Derivation of Eqs.\eqref{eq:TAT},\eqref{eq:TBT},\eqref{eq:TCT},\eqref{eq:WAW},\eqref{eq:WBW} and \eqref{eq:WCW}}
\label{App:ABC}

We present the transformation rules of 
$A_{\ell,\xi}$, $B_{\ell,\xi}$ and 
$C_{\ell,\xi}$ given in Eqs.\eqref{eq:TAT},\eqref{eq:TBT},\eqref{eq:TCT},\eqref{eq:WAW},\eqref{eq:WBW} and \eqref{eq:WCW}. 
Using the assumed form of the Kraus operators 
$\hat{E}_{\ell,\xi}$ defined by 
\eqref{eq:ABC}, we can compute the right hand side of Eq.\eqref{eq:TMT3} as 
\begin{equation}
\hat{T}^\dagger (a) \hat{E}_{\ell,\xi} \hat{T} (a) 
= A_{\ell,\xi}\hat{\mathbb{I}}+\int d^3 p  \,B_{\ell,\xi} (\bm{p}) e^{-ip^\mu a_\mu} \hat{a}(\bm{p})
+\int d^3 p' d^3p \,  C_{\ell,\xi} (\bm{p}', \bm{p}) e^{i({p'}^\mu-p^\mu) a_\mu}\hat{a}^\dagger (\bm{p}') \hat{a}(\bm{p}).
\nonumber
\end{equation}
From Eq.\eqref{eq:TMT3}, we have
\begin{align}
&A_{\ell,\xi}=e^{-i\ell^\mu a_\mu} A_{\ell,\xi}, 
\label{eq:TATA}
\\
&B_{\ell,\xi} (\bm{p}) e^{-ip^\mu a_\mu} =B_{\ell,\xi} (\bm{p}) e^{-i\ell^\mu a_\mu}
\label{eq:TBTA}
\\
&C_{\ell,\xi} (\bm{p}', \bm{p}) e^{i({p'}^\mu-p^\mu) a_\mu}=C_{\ell,\xi} (\bm{p}', \bm{p}) e^{-i\ell^\mu a_\mu}
\label{eq:TCTA}.
\end{align}
The right hand side of Eq.\eqref{eq:WMW} is evaluated as 
\begin{align}
&\hat{V}^\dagger (W) \hat{E}_{\ell,\xi} \hat{V} (W) 
\nonumber 
\\
&
\quad
= A_{\ell,\xi}\hat{\mathbb{I}}+\int d^3 p  \, B_{\ell,\xi} (\bm{p})  \sqrt{\frac{E_{\bm{p}_{W^{-1}}}}{E_{\bm{p}}}}\hat{a}(\bm{p}_{W^{-1}})
+\int d^3 p' d^3p \,  C_{\ell,\xi} (\bm{p}', \bm{p})
\sqrt{\frac{E_{\bm{p}'_{W^{-1}}}}{E_{\bm{p}'}}}
\sqrt{\frac{E_{\bm{p}_{W^{-1}}}}{E_{\bm{p}}}}
\hat{a}^\dagger (\bm{p}'_{W^{-1}}) \hat{a}(\bm{p}_{W^{-1}})
\nonumber
\\
&
\quad
= A_{\ell,\xi}\hat{\mathbb{I}}
+\int d^3 p \, B_{\ell,\xi} (\bm{p}_W)  \sqrt{\frac{E_{\bm{p}_W}}{E_{\bm{p}}}}\hat{a}(\bm{p})
+\int d^3 p' d^3p \, C_{\ell,\xi} (\bm{p}'_W, \bm{p}_W)
\sqrt{\frac{E_{\bm{p}'_W}}{E_{\bm{p}'}}}
\sqrt{\frac{E_{\bm{p}_W}}{E_{\bm{p}}}}
\hat{a}^\dagger (\bm{p}') \hat{a}(\bm{p}),
\nonumber
\end{align}
where note that the Lorentz invariant measure is 
$d^3p/E_{\bm{p}}$ and hence 
$f(\bm{p})d^3p = E_{\bm{p}} f(\bm{p})d^3 p /E_{\bm{p}} = E_{\bm{p}_\Lambda} f(\bm{p}_\Lambda)d^3 p /E_{\bm{p}}$.
From Eq.\eqref{eq:WMW}, we have
\begin{align}
&A_{\ell,\xi}=\sum_{\xi'} \mathcal{D}^*_{\xi'\xi}(W^{-1})  A_{\ell,\xi'}, 
\label{eq:WAWA}
\\
&\sqrt{\frac{E_{\bm{p}_W}}{E_{\bm{p}}}} B_{\ell,\xi} (\bm{p}_W)=\sum_{\xi'}\mathcal{D}^*_{\xi'\xi}(W^{-1}) B_{\ell,\xi'} (\bm{p})
\label{eq:WBWA}
\\
&\sqrt{\frac{E_{\bm{p}'_W}E_{\bm{p}_W}}{E_{\bm{p}'}E_{\bm{p}}}}
 C_{\ell,\xi} (\bm{p}'_W, \bm{p}_W)
=
\sum_{\xi'}\mathcal{D}^*_{\xi'\xi}(W^{-1})
C_{\ell,\xi'} (\bm{p}', \bm{p}) 
\label{eq:WCWA}.
\end{align}

\section{Analysis of a spinless massive particle}
\label{App:massive}

We assume that the spectrum of $\hat{P}^\mu$ on 
any state 
$|\Psi \rangle$ in the Hilbert space of one-particle states, $\mathcal{H}_1$, satisfies 
\begin{equation}
\hat{P}^\mu \hat{P}_\mu |\Psi \rangle =-m^2 |\Psi \rangle, \quad \langle \Psi|\hat{P}^0 |\Psi \rangle >0.
\label{eq:Pmu}
\end{equation}
The above equations are equivalent to the fact that the Hamiltonian 
$\hat{H}=\hat{P}^0$ has the form 
$\hat{H}=\sqrt{\hat{P}_k \hat{P}^k+m^2}$, which implies that 
$|\Psi \rangle$ is the state of a massive particle. 
In this appendix, we derive the form of the dynamical map 
$\mathcal{E}_{t,t_0}$ of a spinless massive particle.

\textbf{\underline{Case I 
 $\, \ell^\mu=[\pm M,0,0,0], \, M>0$ :}}
We focus on the spectrum  
$\ell^\mu=[\pm M,0,0,0], \, M>0$. 
From Eq.\eqref{eq:TATA} for all
$a^\mu=[a,0,0,0]$, 
$A_{\ell,\xi}$ must vanish, 
\begin{equation}
A_{\ell,\xi}=e^{\pm iMa} A_{\ell,\xi} \quad \therefore \quad A_{\ell,\xi}=0.
\nonumber
\end{equation}
Eq.\eqref{eq:TBTA} for all
$a^\mu=[0,\bm{a}]$ leads to
\begin{equation}
B_{\ell,\xi} (\bm{p}) e^{-i\bm{p}\cdot \bm{a}} =B_{\ell,\xi} (\bm{p}) \quad \therefore \quad B_{\ell,\xi} (\bm{p})=B_{\ell,\xi} \delta^3(\bm{p}).
\nonumber
\end{equation}
From Eq.\eqref{eq:TBTA} for all
$a^\mu=[a,0,0,0]$, we get
\begin{equation}
B_{\ell,\xi} (\bm{p}) e^{iE_{\bm{p}} a} =B_{\ell,\xi} (\bm{p}) e^{\pm iMa},
\nonumber
\end{equation}
and combined with 
$B_{\ell,\xi} (\bm{p})=B_{\ell,\xi}  \delta^3(\bm{p})$, we obtain 
\begin{equation}
B_{\ell,\xi}  e^{ima} =B_{\ell,\xi} e^{\pm iMa}. 
\nonumber
\end{equation}
Since the mass 
$m$ is positive, to get a nontrivial result, we should choose 
$+M$ with 
$M=m$.  
Using Eq.\eqref{eq:WBWA} for 
$Q=R \in \text{SO}(3)$ and 
adopting the result 
$B_{\ell,\xi} (\bm{p})=B_{\ell,\xi}  \delta^3(\bm{p})$, we find 
\begin{equation}
B_{\ell,\xi} =\sum_{\xi'}\mathcal{D}^*_{\xi'\xi}(R^{-1}) B_{\ell,\xi'}.
\nonumber
\end{equation}
Since 
$\mathcal{D}_{\xi'\xi}$ is an irreducible (unitary) representation, 
to get a nontrivial 
$B_{\ell,\xi}$, we should choose the spinless representation. 
Hence,
$B_{\ell,\xi}$ is reduced to 
$B_\ell$, which has no leg labeled by 
$\xi$. 
Therefore, 
$B_{\ell} (\bm{p})$, which is given by removing the label 
$\xi$ from 
$B_{\ell,\xi} (\bm{p})$, is
\begin{equation}
 B_{\ell} (\bm{p})=B_\ell \delta^3(\bm{p}).
\nonumber
\end{equation}

From Eq.\eqref{eq:TCTA} for all 
$a^\mu=[0,\bm{a}]$, we deduce
\begin{equation}
C_{\ell,\xi} (\bm{p}',\bm{p}) e^{i(\bm{p}'-\bm{p}) \cdot \bm{a}}=C_{\ell,\xi} (\bm{p}', \bm{p}) \quad \therefore \quad C_{\ell,\xi} (\bm{p}', \bm{p})=C_{\ell,\xi} (\bm{p}) \delta^3(\bm{p}'-\bm{p}).
\nonumber 
\end{equation}
Eq.\eqref{eq:TCTA} for all 
$a^\mu=[a,0,0,0]$ leads to
\begin{equation}
C_{\ell,\xi} (\bm{p}',\bm{p}) e^{-i( E_{\bm{p}'}-E_{\bm{p}} )a}=C_{\ell,\xi} (\bm{p}',\bm{p}) e^{\pm iM a},
\nonumber 
\end{equation}
and substituting 
$C_{\ell,\xi} (\bm{p}',\bm{p})=C_{\ell,\xi} (\bm{p}) \delta^3(\bm{p}'-\bm{p})$ into the above equation, we find that 
$C_{\ell,\xi} (\bm{p})$ vanishes and hence
$C_{\ell,\xi} (\bm{p}',\bm{p})$ is zero,
\begin{equation}
C_{\ell,\xi} (\bm{p})=C_{\ell,\xi} (\bm{p}) e^{\pm iM a} 
\quad \therefore \quad 
C_{\ell,\xi} (\bm{p})=0 
\quad \therefore \quad 
C_{\ell,\xi} (\bm{p}',\bm{p})=0 
\nonumber 
\end{equation}
The above results of 
$A_{\ell,\xi}$, 
$B_{\ell,\xi}(\bm{p})$ and 
$C_{\ell,\xi} (\bm{p}',\bm{p})$
give the Kraus operator 
$\hat{E}_{\ell,\xi}$ with 
$\ell^\mu=[m,0,0,0]$ as 
\begin{equation}
\hat{E}_{\ell,\xi}=  B_\ell\hat{a}(\bm{0}) .
\nonumber 
\end{equation}
Eq.\eqref{eq:SMS} tells us that 
\begin{equation}
\hat{E}_{q,\xi}=  N^*_q \hat{V}(S_q) \hat{E}_{\ell,\xi} \hat{V}^\dagger(S_q) =N^*_q B_\ell \sqrt{\frac{E_{\bm{q}}}{m}} \hat{a}(\bm{q}),
\nonumber
\end{equation}
where 
$E_q=(S_q \, \ell)^0$ and 
$q^i=(S_q \, \ell)^i$. 
Choosing the normalization of the inner product 
$\bm{v}^\dagger_{q',\xi'} \bm{v}_{q,\xi}$ as
\begin{equation}
\bm{v}^\dagger_{q',t_0'} \bm{v}_{q,\xi}=\delta^3(\bm{q}'-\bm{q}) \delta_{\xi'\xi},
\nonumber
\end{equation}
we have 
$N_q=\sqrt{m/E_{\bm{q}}}$ up to a phase factor and the completeness condition,
\begin{equation}
\int d^3 q \sum_{s} \bm{v}_{q,\xi} \bm{v}^\dagger_{q,\xi}=\bm{I}.
\nonumber 
\end{equation}
We then derive a part of the dynamical map $\mathcal{E}_{t,t_0}$ as 
\begin{equation}
\mathcal{E}_{t,t_0}[\rho(t_0)] 
\supset |B_\ell|^2 \, \int d^3 q \, \hat{a}(\bm{q}) \rho(t_0) \hat{a}^\dagger (\bm{q}).
\label{eq:EI}
\end{equation}

\underline{\textbf{Case II  
$\, \ell^\mu=[\pm\kappa,0,0,\kappa], \, \kappa>0$ :}}
We consider the spectrum 
$\ell^\mu=[\pm \kappa,0,0,\kappa], \, \kappa>0$. 
From Eq.\eqref{eq:TATA} for all
$a^\mu=[a,0,0,0]$, 
$A_{\ell,\xi}$ turns out to be zero,
\begin{equation}
A_{\ell,\xi}=e^{\pm i\kappa a} A_{\ell,\xi} \quad \therefore \quad A_{\ell,\xi}=0.
\nonumber
\end{equation}

From Eq.\eqref{eq:TBTA} for all
$a^\mu=[0,\bm{a}]$, we get
\begin{equation}
B_{\ell,\xi} (\bm{p}) e^{-i\bm{p} \cdot \bm{a}} =B_{\ell,\xi} (\bm{p})e^{-i\bm{\ell} \cdot \bm{a}} 
\quad \therefore \quad 
B_{\ell,\xi} (\bm{p})=B_{\ell,\xi}  \delta^3(\bm{p}-\bm{\ell}),
\nonumber
\end{equation}
where 
$\bm{\ell} =[0,0,\kappa]^\text{T}$. 
Eq.\eqref{eq:TBTA} for all 
$a^\mu=[a,0,0,0]$ leads to 
\begin{equation}
B_{\ell,\xi} (\bm{p}) e^{iE_{\bm{p}} a} =B_{\ell,\xi} (\bm{p}) e^{\pm i\kappa a},
\nonumber
\end{equation}
and substituting 
$B_{\ell,\xi} (\bm{p})=B_{\ell,\xi}  \delta^3(\bm{p}-\bm{\ell})$ and $E_{\bm{\ell}}=\sqrt{\bm{\ell}^2+m^2}=\sqrt{\kappa^2+m^2}$ into the above equation, we find that 
$B_{\ell,\xi}$ and 
$B_{\ell,\xi} (\bm{p})$ are trivial, 
\begin{equation}
B_{\ell,\xi} e^{i\sqrt{\kappa^2 +m^2}a}=B_{\ell,\xi} e^{\pm i\kappa a} 
\quad \therefore \quad 
B_{\ell,\xi}=0 
\quad \therefore \quad
B_{\ell,\xi} (\bm{p})=0.
\nonumber
\end{equation}

Eq.\eqref{eq:TCTA} for all 
$a^\mu=[0,\bm{a}]$ gives
\begin{equation}
C_{\ell,\xi} (\bm{p}',\bm{p}) e^{i(\bm{p}'-\bm{p}) \cdot \bm{a}}=C_{\ell,\xi} (\bm{p}',\bm{p})e^{-i\bm{\ell}\cdot \bm{a}} \quad \therefore \quad C_{\ell,\xi} (\bm{p}',\bm{p})=C_{\ell,\xi} (\bm{p}) \delta^3(\bm{p}'-\bm{p}+\bm{\ell}).
\nonumber 
\end{equation}
Using Eq.\eqref{eq:TCTA} for all 
$a^\mu=[a,0,0,0]$, we get
\begin{equation}
C_{\ell,\xi} (\bm{p}',\bm{p}) e^{-i( E_{\bm{p}'}-E_{\bm{p}} )a}=C_{\ell,\xi} (\bm{p}',\bm{p}) e^{\pm i\kappa a},
\nonumber 
\end{equation}
and substituting 
$C_{\ell,\xi} (\bm{p}',\bm{p})=C_{\ell,\xi} (\bm{p}) \delta^3(\bm{p}'-\bm{p}+\bm{\ell})$ into the above equation, we have
\begin{equation}
C_{\ell,\xi} (\bm{p})e^{-i( E_{\bm{p}-\bm{\ell}}-E_{\bm{p}} )a}=C_{\ell,\xi} (\bm{p}) e^{\pm i\kappa a}.
\nonumber 
\end{equation}
Noticing the fact that 
$E_{\bm{p}-\bm{\ell}}-E_{\bm{p}} \pm \kappa \neq 0$, we get  
$C_{\ell,\xi} (\bm{p})=0$ and 
\begin{equation}
C_{\ell,\xi} (\bm{p}',\bm{p})=0.
\nonumber 
\end{equation}

Combined with the above analyses on 
$A_{\ell,\xi}$, $B_{\ell,\xi}(\bm{p})$ and 
$C_{\ell,\xi} (\bm{p}',\bm{p})$, the Kraus operator 
$\hat{E}_{\ell,\xi}$ vanishes:
\begin{equation}
\hat{E}_{\ell,\xi}=0 
\quad \therefore \quad 
\hat{E}_{q,\xi}=N_q \hat{V}(S_q) \hat{E}_{\ell,\xi} \hat{V}^\dagger (S_q)=0.
\label{eq:EII}
\end{equation}

\underline{\textbf{Case III 
$\,\ell^\mu=[0,0,0,w], \, N^2>0$ :}}
We focus on the spectrum 
$\ell^\mu=[0,0,0,w], \, N^2>0$. 
From Eq.\eqref{eq:TATA} for all
$a^\mu=[0,\bm{a}]$, we have
\begin{equation}
A_{\ell,\xi}=e^{-i\bm{\ell} \cdot \bm{a} } A_{\ell,\xi} \quad \therefore \quad A_{\ell,\xi}=0.
\nonumber
\end{equation}

Eq.\eqref{eq:TBTA} for all 
$a^\mu=[a,0,0,0]$ leads to 
\begin{equation}
B_{\ell,\xi} (\bm{p}) e^{iE_{\bm{p}} a} =B_{\ell,\xi} (\bm{p}) 
\quad \therefore \quad 
B_{\ell,\xi} (\bm{p})=0,
\nonumber
\end{equation}
where note that 
$E_{\bm{q}}=\sqrt{\bm{q}^2+m^2} \neq 0$. 

From Eq.\eqref{eq:TCTA} for all 
$a^\mu=[0,\bm{a}]$, we deduce
\begin{equation}
C_{\ell,\xi} (\bm{p}',\bm{p}) e^{i(\bm{p}'-\bm{p}) \cdot \bm{a}}=C_{\ell,\xi} (\bm{p}',\bm{p})e^{-i\bm{\ell}\cdot \bm{a}} \quad \therefore \quad C_{\ell,\xi} (\bm{p}',\bm{p})=C_{\ell,\xi} (\bm{p}) \delta^3(\bm{p}'-\bm{p}+\bm{\ell}),
\nonumber 
\end{equation}
where 
$\bm{\ell}=[0,0,w]^\text{T}$. 
From Eq.\eqref{eq:TCTA} for all 
$a^\mu=[a,0,0,0]$, we get
\begin{equation}
C_{\ell,\xi} (\bm{p}',\bm{p}) e^{-i( E_{\bm{p}'}-E_{\bm{p}} )a}=C_{\ell,\xi} (\bm{p}',\bm{p}),
\nonumber 
\end{equation}
and substituting 
$C_{\ell,\xi} (\bm{p}',\bm{p})=C_{\ell,\xi} (\bm{p}) \delta^3(\bm{p}'-\bm{p}+\bm{\ell})$ into the above equation, we have
\begin{equation}
C_{\ell,\xi} (\bm{p})e^{-i( E_{\bm{p}-\bm{\ell}}-E_{\bm{p}} )a}=C_{\ell,\xi} (\bm{p}) 
\quad \therefore \quad 
C_{\ell,\xi} (\bm{p})=C_{\ell,\xi} (\bm{p}_\perp) \delta(p^z-N/2).
\nonumber 
\end{equation}
Substituting this into
$C_{\ell,\xi} (\bm{p}',\bm{p})=C_{\ell,\xi} (\bm{p}) \delta^3(\bm{p}'-\bm{p}+\bm{\ell})$, we obtain
\begin{equation}
C_{\ell,\xi} (\bm{p}',\bm{p})=C_{\ell,\xi} (\bm{p}_\perp) \delta^2(\bm{p}'_\perp-\bm{p}_\perp) \delta(p'^z+N/2) \delta(p^z-N/2). 
\nonumber 
\end{equation}

With the above results of 
$A_{\ell,\xi}$, 
$B_{\ell,\xi}(\bm{p})$ and 
$C_{\ell,\xi} (\bm{p}',\bm{p})$,
the Kraus operator 
$\hat{E}_{\ell,\xi}$ is written as 
\begin{equation}
\hat{E}_{\ell,\xi}= \int d^2 p_\perp  C_{\ell,\xi} (\bm{p}_\perp) \hat{a}^\dagger (\bm{p}_\perp,-N/2) \hat{a}(\bm{p}_\perp,N/2).
\nonumber 
\end{equation}
By the completeness condition of the Kraus operators, Eq.\eqref{eq:EdgE}, the above Kraus operator 
$\hat{E}_{\ell,\xi}$ should satisfy 
$\hat{E}^\dagger_{\ell,\xi} \hat{E}_{\ell,\xi} \leq \hat{\mathbb{I}}$. 
Concretely, 
$\hat{E}^\dagger_{\ell,\xi} \hat{E}_{\ell,\xi} $ is evaluated as 
\begin{align}
\hat{E}_{\ell,\xi}^\dagger \hat{E}_{\ell,\xi}
&=\int d^2 p'_\perp   C^*_{\ell,\xi} (\bm{p}'_\perp) \hat{a}^\dagger (\bm{p}'_\perp,N/2)\hat{a} (\bm{p}'_\perp,-N/2)
\int d^2 p_\perp C_{\ell,\xi} (\bm{p}_\perp) \hat{a}^\dagger (\bm{p}_\perp,-N/2) \hat{a}(\bm{p}_\perp,N/2)
\nonumber 
\\
&=\delta(0) 
\int d^2 p_\perp
 \Big[C_{\ell,\xi} (\bm{p}_\perp) \hat{a} (\bm{p}_\perp,N/2) \Big]^\dagger \Big[ C_{\ell,\xi} (\bm{p}_\perp)  \hat{a}(\bm{p}_\perp,N/2)\Big],
\nonumber 
\end{align}
where the term given by the linear combination of 
$\hat{a}^\dagger \hat{a}^\dagger \hat{a}  \hat{a}$ vanishes on 
$\mathcal{H}_0 \oplus \mathcal{H}_1$. 
To satisfy 
$\hat{E}^\dagger_{\ell,\xi} \hat{E}_{\ell,\xi} \leq \hat{\mathbb{I}}$, we find that 
\begin{align}
\int d^2 p_\perp \Big[ C_{\ell,\xi} (\bm{p}_\perp) \hat{a} (\bm{p}_\perp,N/2) \Big]^\dagger \Big[ C_{\ell,\xi} (\bm{p}_\perp)  \hat{a}(\bm{p}_\perp,N/2)\Big]=0,
\nonumber 
\end{align}
which leads to 
$C_{\ell,\xi} (\bm{p}_\perp)=0$ and 
$\langle \Phi|\hat{E}_{\ell,\xi}|\Psi \rangle=0$ for all 
$|\Psi \rangle, |\Phi \rangle \in \mathcal{H}_0 \oplus \mathcal{H}_1$. 
Hence, the Kraus operator 
$\hat{E}_{q,\xi}$ vanishes,
\begin{equation}
\hat{E}_{q,\xi}
=N^*_q \hat{V}(S_q) \hat{E}_{\ell,\xi} \hat{V}^\dagger (S_q)=0,
\label{eq:EIII}
\end{equation}
on the Hilbert space $\mathcal{H}_0 \oplus \mathcal{H}_1$.

\underline{\textbf{Case IV
$\ell^\mu=[0,0,0,0]$ :}}
We consider the case where 
$\ell^\mu=[0,0,0,0]$. 
Before computations, we mention about the unitary irreducible representations 
$\mathcal{D}(\Lambda)$ of the Lorentz group \cite{Bargmann1947, Tung1985}.
Let 
$\bm{\mathcal{J}}$ and
$\bm{\mathcal{K}}$ be the generators of $\mathcal{D}(\Lambda)$, which give rotations and boosts, respectively.  
The unitary irreducible representation 
$\mathcal{D}(\Lambda)$ is classified by two parameters 
$j_0$ and 
$\nu$, by which the eigenvalues of the two Casimir operators 
$I_1=\bm{\mathcal{J}}^2-\bm{\mathcal{
K}}^2$ and 
$I_2=\bm{\mathcal{J}} \cdot \bm{\mathcal{K}}$ are determined. 
The value 
$j_0(j_0 +1)$ gives the minimum eigenvalue of 
$\bm{\mathcal{J}}^2$. 
The parameters 
$j_0$ and 
$\nu$ are decomposed into two parts called (1) the principal series 
$-i \nu \in \mathbb{R}$ and 
$j_0=0, 1/2, 1, \dots$ and (2) the complementary series 
$0 \leq \nu^2 \leq 1$ and 
$j_0=0$.  
We denote 
$\mathcal{D}^{j_0,\nu} (\Lambda)$ as the unitary irreducible representation with 
$j_0$ and 
$\nu$. 
In particular,   
$\mathcal{D}^{0,\pm 1}(\Lambda)$ is the trivial one-dimensional representation $\mathcal{D}^{0,1}(\Lambda)=\mathcal{D}^{0,-1}(\Lambda)=1$ and others are the infinite-dimensional representations.

With the above knowledge on the unitary irreducible representation of the Lorentz group, we proceed with computations. 
In the following, we drop the label 
$\ell$.  
Eq.\eqref{eq:TATA} is identical for all 
$a^\mu$. 
Since the little group associated with 
$\ell^\mu$ is 
$\text{SO}(3,1)$,
Eq.\eqref{eq:WAWA} for 
$W=\Lambda \in \text{SO}(3,1)$ is given as 
\begin{equation}
A_{\xi} =\sum_{\xi'}\mathcal{D}^*_{\xi'\xi}(\Lambda^{-1})A_{\xi'}.
\label{eq:Axi(f)}
\end{equation}
For the one-dimensional representation, since 
the representation is trivial 
$\mathcal{D}^{0,1}(\Lambda)=\mathcal{D}^{0,-1}(\Lambda)=1$,  
$A_\xi$ is a scalar,
$A_\xi=A$. 
Then, Eq.\eqref{eq:Axi(f)} trivially holds. 
For the infinite-dimensional representation, 
$A_\xi=0$ from Eq.\eqref{eq:Axi(f)}.

Eq.\eqref{eq:TBTA} for all 
$a^\mu=[a,0,0,0]$ gives the condition 
\begin{equation}
B_{\xi} (\bm{p}) e^{iE_{\bm{p}} a} =B_{\xi} (\bm{p}) 
\quad \therefore \quad 
B_{\xi} (\bm{p})=0,
\nonumber
\end{equation}
where note that 
$E_{\bm{p}}=\sqrt{\bm{p}^2+m^2} \neq 0$. 

From Eq.\eqref{eq:TCTA} for all 
$a^\mu$, we obtain
\begin{equation}
C_{\xi} (\bm{p}', \bm{p}) e^{i({p'}^\mu-p^\mu) a_\mu }=C_{\xi} (\bm{p}', \bm{p}) \quad \therefore \quad 
C_{\xi} (\bm{p}',\bm{p})=C_{\xi} (\bm{p}) \delta^3(\bm{p}'-\bm{p}).
\nonumber 
\end{equation}
Eq. \eqref{eq:WCWA} for 
$W=\Lambda \in \text{SO}(3,1)$ is written as 
\begin{equation}
\sqrt{\frac{E_{\bm{p}'_\Lambda}E_{\bm{p}_\Lambda}}{E_{\bm{p}' }E_{\bm{p}}}}
 C_{\xi} (\bm{p}'_\Lambda, \bm{p}_\Lambda)
=
\sum_{\xi'}\mathcal{D}^*_{\xi'\xi}(\Lambda^{-1})
C_{\xi'} (\bm{p}', \bm{p}).
\nonumber
\end{equation}
The equation 
$C_{\xi} (\bm{p}',\bm{p})=C_{\xi} (\bm{p}) \delta^3(\bm{p}'-\bm{p})$ and the fact that the invariant delta function is 
$E_{\bm{p}} \delta^3(\bm{p}-\bm{p}')$ yield
\begin{equation}
 C_{\xi} (\bm{p}_\Lambda)
=
\sum_{\xi'}\mathcal{D}^*_{\xi'\xi}(\Lambda^{-1})
C_{\xi'} (\bm{p}).
\label{eq:Cxi(f)}
\end{equation}
Choosing 
$\bm{p}=\bm{0}$ and 
$\Lambda=S_q$ with 
$(S_q)^\mu{}_\nu k^\nu=q^\mu$ for 
$k^\mu=[m,0,0,0]$, we have
\begin{equation}
 C_\xi (\bm{q})
=
\sum_{\xi'}\mathcal{D}^*_{\xi'\xi}(S_q^{-1})
C_{\xi'} (\bm{0}).
\label{eq:CSq} 
\end{equation}
Using Eq.\eqref{eq:CSq}, Eq.\eqref{eq:Cxi(f)} is written as 
\begin{equation}
 C_{\xi} (\bm{0})=
\sum_{\xi'}\mathcal{D}^*_{\xi'\xi}(Q)
C_{\xi'} (\bm{0}),
\label{eq:Cxi(f)'}
\end{equation}
where 
$\sum_{\xi''} \mathcal{D}_{\xi\xi''}(\Lambda)\mathcal{D}_{\xi'' \xi'}(\Lambda')=\mathcal{D}_{\xi\xi'}(\Lambda \Lambda')$ and 
$\mathcal{D}^*_{\xi' \xi}(\Lambda)=\mathcal{D}_{\xi\xi'}(\Lambda^{-1})$ were used, and 
$Q=Q(\Lambda,p)=S^{-1}_{\Lambda p} \Lambda S_p \in \text{SO}(3)$.
For the trivial representation 
$\mathcal{D}(\Lambda)=\mathcal{D}^{0,\pm 1}(\Lambda)=1$, the label 
$\xi$ is removed and we should use
$C (\bm{p})$ instead of 
$C_\xi (\bm{p})$.
Eq.\eqref{eq:Cxi(f)'} is then automatically satisfied.  
From Eq.\eqref{eq:CSq}, we find that 
$C(\bm{q})$ is just a constant 
$C$:
\begin{equation}
 C(\bm{q})
=
C(\bm{0})=C.
\nonumber 
\end{equation}

For the infinite-dimensional representation  
$\mathcal{D}(\Lambda)=\mathcal{D}^{j_0,\nu}(\Lambda)$ with $(j_0,\nu) \neq (0,\pm 1)$, we have 
$C_{\xi} (\bm{p}',\bm{p})=C_{\xi;\,j_0,\nu} (\bm{p}', \bm{p})$ with
\begin{equation}
C_{\xi;\,j_0,\nu} (\bm{p}', \bm{p})=C_{\xi;\,j_0,\nu} (\bm{p}) \delta^3(\bm{p}'-\bm{p}),
\nonumber 
\end{equation}
where the dependence of 
$j_0$ and 
$\nu$ was explicitly indicated. 
For a spinless massive particle, the equation 
\begin{equation}
C_{\xi;\,j_0,\nu} (\bm{0})
=
\sum_{\xi'}\mathcal{D}^{j_0,\nu \,*}_{\xi'\xi}(Q)
C_{\xi';\,j_0,\nu} (\bm{0}) 
\nonumber
\end{equation}
is satisfied from Eq.\eqref{eq:Cxi(f)'}, where 
$Q \in \text{SO}(3)$. 
To get a nontrivial solution, 
$j_0$ must be zero, and 
$C_{\xi;\,0,\nu} (\bm{0})$ belongs to the representation space with $j_0=0$. 
Hence, 
$C_{\xi;\,0,\nu} (\bm{0})=C_\nu \delta_{\xi\xi_0}$ with 
$\mathcal{D}^{0,\nu}_{\xi \xi_0} (Q)=\delta_{\xi \xi_0}$ for 
$Q \in \text{SO}(3)$. 
Eq.\eqref{eq:CSq} gives 
\begin{equation}
C_{\xi;\,0,\nu} (\bm{q})
=C_\nu \mathcal{D}^{0,\nu \,*}_{\xi_0 \xi}(S^{-1}_q),
\label{eq:Cxi}
\end{equation}
where note that 
$S_q$ is not an element of 
$\text{SO}(3)$. 

The above analysis for the case 
$\ell^\mu=[0,0,0,0]$ tells us the following Kraus operators
\begin{equation}
\hat{E} = A \hat{\mathbb{I}}+C \hat{N},
\quad 
\hat{E}_{\xi;0,\nu} = \int d^3p \, C_\nu \mathcal{D}^{0,\nu \,*}_{\xi_0 \xi}(S^{-1}_p) \hat{a}^\dagger (\bm{p}) \hat{a}(\bm{p}),
\nonumber 
\end{equation}
where 
$A$ and 
$C$ is a complex number, 
$\hat{N}$ is the number operator defined in \eqref{eq:N}. 
A part of the dynamical map 
$\mathcal{E}_{t,t_0}$ is given as  
\begin{align}
\mathcal{E}_{t,t_0}[\rho(t_0)] 
&\supset 
\Big(
A \hat{\mathbb{I}}+C \hat{N} 
\Big) 
\rho(t_0) \Big(
A \hat{\mathbb{I}}+C \hat{N}
\Big)^\dagger + \sum_{\xi,\nu} \hat{E}_{\xi;0, \nu} \rho(t_0) \hat{E}^\dagger_{\xi; 0, \nu}
\nonumber 
\\ 
&= 
\Big(
A \hat{\mathbb{I}}+C \hat{N} 
\Big) 
\rho(t_0) \Big(
A \hat{\mathbb{I}}+C \hat{N}
\Big)^\dagger + \int d^3p d^3q D(p,q) \hat{a}^\dagger (\bm{p}) \hat{a}(\bm{p}) \rho(t_0) \hat{a}^\dagger (\bm{q}) \hat{a}(\bm{q}) 
\label{eq:EIV}
\end{align}
where 
$\sum_{\nu}$ is defined by
\begin{equation}
\sum_{\nu} F_{\nu}= \int_{-1< \nu < 1} d\nu F_{\nu}+\int_{-i\nu \in \mathbb{R}} d(-i\nu)   F_{\nu},
\nonumber 
\end{equation}
and 
$D(p,q)$ is 
\begin{equation}
D(p,q)=\sum_{\nu,\xi}|C_\nu|^2 \mathcal{D}^{0,\nu \,*}_{\xi_0 \xi}(S^{-1}_p) \mathcal{D}^{0,\nu }_{\xi_0 \xi}(S^{-1}_q).
\label{eq:D}
\end{equation}
This function 
$D(p,q)$ is non-negative and Lorentz invariant in the sense that 
\begin{equation}
\int d^3p d^3q f^*(p)D(p,q) f(q) \geq 0, \quad D(\Lambda p, \Lambda q) = D(p,q), 
\label{eq:Dcond}
\end{equation}
where 
$f(p)$ is a complex function of the particle's four momentum and 
$\Lambda$ is the Lorentz transformaiton matrix. 
It is easy to show the former condition, and the latter equation is shown as 
\begin{align}
D(\Lambda p, \Lambda q)
&=\sum_{\nu,\xi}|C_\nu|^2 \mathcal{D}^{0,\nu \,*}_{\xi_0 \xi}(S^{-1}_{\Lambda p}) \mathcal{D}^{0,\nu }_{\xi_0 \xi}(S^{-1}_{\Lambda q}) 
\nonumber 
\\
&=\sum_{\nu,\xi}|C_\nu|^2 \mathcal{D}^{0,\nu \,*}_{\xi_0 \xi}(Q(\Lambda,p) S^{-1}_p \Lambda^{-1}) \mathcal{D}^{0,\nu }_{\xi_0 \xi}(Q(\Lambda,q) S^{-1}_q \Lambda^{-1}) 
\nonumber 
\\
&=\sum_{\nu,\xi}|C_\nu|^2 
\sum_{\xi'} \mathcal{D}^{0,\nu \,*}_{\xi_0 \xi'}(Q(\Lambda,p) )  \mathcal{D}^{0,\nu \,*}_{\xi' \xi}(S^{-1}_p \Lambda^{-1}) 
\sum_{\xi''}
\mathcal{D}^{0,\nu }_{\xi_0 \xi''}(Q(\Lambda,q) ) \mathcal{D}^{0,\nu }_{\xi'' \xi}( S^{-1}_q \Lambda^{-1}) 
\nonumber 
\\
&=\sum_{\nu,\xi}|C_\nu|^2 
\mathcal{D}^{0,\nu \,*}_{\xi_0 \xi}(S^{-1}_p \Lambda^{-1}) \mathcal{D}^{0,\nu }_{\xi_0 \xi}( S^{-1}_q \Lambda^{-1}) 
\nonumber 
\\
&=\sum_{\nu,\xi}|C_\nu|^2 
\sum_{\xi'} \mathcal{D}^{0,\nu \,*}_{\xi_0 \xi'}(S^{-1}_p ) \mathcal{D}^{0,\nu \,*}_{\xi' \xi}(\Lambda^{-1}) 
\sum_{\xi''} \mathcal{D}^{0,\nu }_{\xi_0 \xi''}( S^{-1}_q ) \mathcal{D}^{0,\nu }_{\xi'' \xi}( \Lambda^{-1}) 
\nonumber 
\\
&=\sum_{\nu}|C_\nu|^2 
\sum_{\xi', \xi''} \mathcal{D}^{0,\nu \,*}_{\xi_0 \xi'}(S^{-1}_p ) \delta_{\xi'\xi''}
\mathcal{D}^{0,\nu }_{\xi_0 \xi''}( S^{-1}_q )  
\nonumber 
\\
&=\sum_{\nu,\xi}|C_\nu|^2 
\mathcal{D}^{0,\nu \,*}_{\xi_0 \xi}(S^{-1}_p ) \mathcal{D}^{0,\nu }_{\xi_0 \xi}( S^{-1}_q )
\nonumber 
\\
&=D(p,q),
\nonumber
\end{align}
where we used the equations 
$Q(\Lambda,p)=S^{-1}_{\Lambda p} \Lambda p$, 
$\sum_{\xi''} \mathcal{D}^{0,\nu}_{\xi\xi''}(\Lambda)\mathcal{D}^{0,\nu}_{\xi'' \xi'}(\Lambda')=\mathcal{D}^{0,\nu}_{\xi\xi'}(\Lambda \Lambda')$,
$\mathcal{D}^{0,\nu \,*}_{\xi' \xi}(\Lambda)=\mathcal{D}^{0,\nu}_{\xi\xi'}(\Lambda^{-1})$ and $\mathcal{D}^{0,\nu }_{\xi_0 \xi}(Q)=\delta_{\xi_0 \xi}$ for 
$Q \in \text{SO}(3)$. 

\underline{\textbf{Summary from Case I to Case IV:}} Let us summarize the analysis of a spinless massive particle. 
The above results given in Eqs.\eqref{eq:EI}, \eqref{eq:EII}, \eqref{eq:EIII} and \eqref{eq:EIV} provide the following form of 
$\mathcal{E}_{t,t_0}$: 
\begin{align}
\mathcal{E}_{t,t_0}[\rho(t_0)] 
&= |B|^2 \, \int d^3 p \, \hat{a}(\bm{p}) \rho(t_0) \hat{a}^\dagger (\bm{p})+ \Big(
A \hat{\mathbb{I}}+C \hat{N} 
\Big) 
\rho(t_0) \Big(
A \hat{\mathbb{I}}+C \hat{N}
\Big)^\dagger
\nonumber 
\\
&
+ \int d^3p d^3q D(p,q) \hat{a}^\dagger (\bm{p}) \hat{a}(\bm{p}) \rho(t_0) \hat{a}^\dagger (\bm{q}) \hat{a}(\bm{q}),
\nonumber
\end{align}
where 
$B_\ell$ is denoted by 
$B$ for simplicity. 
For the Hilbert space 
$\mathcal{H}_0 \oplus \mathcal{H}_1$ of the vacuum state and one-particle states of the massive particle, the completeness condition of the Kraus operators gives 
\begin{align}
\hat{\mathbb{I}}
&=
|B|^2 \int d^3 p \hat{a}^\dagger (\bm{p})\hat{a}(\bm{p}) + \Big(
\hat{A\mathbb{I}}+C \hat{N} \Big)^\dagger\Big(
\hat{A\mathbb{I}}+C \hat{N} \Big)
+\int d^3p d^3q D (p,q) \hat{a}^\dagger (\bm{q}) \hat{a}(\bm{q}) \hat{a}^\dagger (\bm{p}) \hat{a}(\bm{p}) 
\nonumber 
\\
&=
|A|^2
\hat{\mathbb{I}}+ \big[|B|^2+ AC^*+A^*C+|C|^2 +D \big] \hat{N} ,
\label{eq:comp}
\end{align}
where note that the equations 
\begin{equation}
\hat{N}^2=\hat{N}, 
\quad
\hat{a}^\dagger (\bm{q})\hat{a}(\bm{q})\hat{a}^\dagger(\bm{p})\hat{a}(\bm{p})
=\hat{a}^\dagger (\bm{p})\hat{a}(\bm{p}) \delta^3(\bm{p}-\bm{q}).
\label{eq:N2=N}
\end{equation}
hold on $\mathcal{H}_0 \oplus \mathcal{H}_1$ and that 
$D=D(p,p)$ does not depend on the three-momentum 
$\bm{p}$ due to the Lorentz invariance (the latter equation in \eqref{eq:Dcond}).
Eq.\eqref{eq:comp} gives
\begin{align}
|A|^2 =1, 
\quad 
|B|^2+AC^*+A^*C+|C|^2 +D=0.
\label{eq:comp2}
\end{align}

The parameters
$A$, $B$, 
$C$ and the function 
$D(p,q)$ may depend on 
$t$ and 
$t_0$.
Redefining  
$|B|^2$, 
$C/A$, and 
$D(p,q)$ as 
$\beta_{t,t_0} $, 
$\gamma_{t,t_0}$, and 
$\delta_{t,t_0} (p,q)$, respectively, we get the following dynamical map 
$\mathcal{E}_{t,t_0}$,
\begin{align}
\mathcal{E}_{t,t_0}[\rho(t_0)] 
&= \beta_{t,t_0}  \int d^3 p \hat{a}(\bm{p}) \rho(t_0) \hat{a}^\dagger (\bm{p})+ \Big(
 \hat{\mathbb{I}}+\gamma_{t,t_0} \hat{N} 
\Big) 
\rho(t_0) \Big(
 \hat{\mathbb{I}}+\gamma_{t,t_0} \hat{N}
\Big)^\dagger
\nonumber 
\\
&+ \int d^3p d^3q \delta_{t,t_0} (p,q) \hat{a}^\dagger (\bm{p}) \hat{a}(\bm{p}) \rho(t_0) \hat{a}^\dagger (\bm{q}) \hat{a}(\bm{q}).
\label{eq:Emassive}
\end{align}
This is nothing but \eqref{eq:map}. 
According to the definitions 
$\beta_{t,t_0} $, 
$\gamma_{t,t_0}$, and 
$\delta_{t,t_0} (p,q)$, we find that the function 
$\delta_{t,t_0} (p,q)$
is non-negative and Lorentz invariant,
\begin{equation}
\int d^3p d^3q f^*(p) \delta_{t,t_0} (p,q) f(q) \geq 0, \quad \delta_{t,t_0}(\Lambda p, \Lambda q) = \delta_{t,t_0}(p,q),
\label{eq:deltacond}
\end{equation}
and that the parameters
$\beta_{t,t_0} $, 
$\gamma_{t,t_0}$ and
$\delta_{t,t_0}=\delta_{t,t_0} (p,p) $
satisfy 
\begin{equation}
\beta_{t,t_0} \geq 0, \quad \beta_{t,t_0}+\gamma^*_{t,t_0}+\gamma_{t,t_0}+|\gamma_{t,t_0}|^2 +\delta_{t,t_0}=0.
\label{eq:comp3}
\end{equation}

\section{Computation of the characteristic function}
\label{App:consv}

In this appendix, we derive Eqs.\eqref{eq:chit2} and \eqref{eq:chit3}. 
Since 
$\rho(t)=\Phi_{t,t_0}[\rho(t_0)]=\mathcal{U}_{t,t_0} \circ \mathcal{E}_{t,t_0}[\rho(t_0)]$ and 
$\mathcal{U}_{t,t_0}
[\rho]=e^{-i\hat{H}(t-t_0)} \rho e^{i\hat{H}(t-t_0)}$, 
we have 
\begin{align}
\chi_t (0,a)
=\text{Tr}[e^{-ia_\mu \hat{P}^\mu } \rho(t)]
=\text{Tr}[e^{-ia_\mu \hat{P}^\mu } \mathcal{U}_{t,t_0} \circ \mathcal{E}_{t,t_0}[\rho(t_0)]]
=\text{Tr}[e^{-ia_\mu \hat{P}^\mu } \mathcal{E}_{t,t_0}[\rho(t_0)]],
\nonumber
\end{align}
where 
$[\hat{H},\hat{P}^\mu]=0$ was used.
Substituting the form of 
$\mathcal{E}_{t,t_0}$ given in \eqref{eq:map} into the above equation, we get 
\begin{align}
\chi_t (0,a)
&=\beta_{t,t_0} \text{Tr}\Big[e^{-ia_\mu \hat{P}^\mu } \int d^3p\,\hat{a}(\bm{p}) \rho(t_0)\hat{a}^\dagger(\bm{p})\Big]
+ \text{Tr}[e^{-ia_\mu \hat{P}^\mu } (\hat{\mathbb{I}}+\gamma_{t,t_0} \hat{N})\rho(t_0)(\hat{\mathbb{I}}+\gamma^*_{t,t_0 }\hat{N})]
\nonumber 
\\
&+
 \text{Tr}\Big[e^{-ia_\mu \hat{P}^\mu } \int d^3p \int d^3q \, \delta_{t,t_0} (p,q)
\hat{a}^\dagger(\bm{p})\hat{a}(\bm{p}) 
\, \rho(t_0) \, 
\hat{a}^\dagger (\bm{q})\hat{a}(\bm{q}) \Big].
\nonumber
\end{align}
Since 
$e^{-i\hat{P}^\mu a_\mu}\hat{N}e^{i\hat{P}^\mu a_\mu}=\hat{N}$ and 
$e^{-i\hat{P}^\mu a_\mu}\hat{a}^\dagger(\bm{p})\hat{a}(\bm{p}) e^{i\hat{P}^\mu a_\mu}=\hat{a}^\dagger(\bm{p})\hat{a}(\bm{p}) $, the characteristic function is written as 
\begin{align}
\chi_t (0,a)
&=\beta_{t,t_0} \text{Tr}\Big[e^{-ia_\mu \hat{P}^\mu } \int d^3p\,\hat{a}(\bm{p}) \rho(t_0)\hat{a}^\dagger(\bm{p})\Big]
+ \text{Tr}[ (\hat{\mathbb{I}}+\gamma^*_{t,t_0 }\hat{N})(\hat{\mathbb{I}}+\gamma_{t,t_0} \hat{N})e^{-ia_\mu \hat{P}^\mu }\rho(t_0)]
\nonumber 
\\
&+
 \text{Tr}\Big[ \int d^3p \int d^3q \, \delta_{t,t_0} (p,q)
\hat{a}^\dagger (\bm{q})\hat{a}(\bm{q})\hat{a}^\dagger(\bm{p})\hat{a}(\bm{p}) 
\, e^{-ia_\mu \hat{P}^\mu }\rho(t_0) \Big].
\nonumber
\end{align}
On the Hilbert space
$\mathcal{H}_0 \oplus \mathcal{H}_1$ of the vacuum state and one-particle states of the massive particle, the following equations hold:
\begin{equation}
e^{-i\hat{P}_\mu a^\mu} \hat{a}(\bm{p})=\hat{a
}(\bm{p}), 
\quad \hat{N}^2=\hat{N}, 
\quad
\hat{a}^\dagger (\bm{q})\hat{a}(\bm{q})\hat{a}^\dagger(\bm{p})\hat{a}(\bm{p})
=\hat{a}^\dagger (\bm{p})\hat{a}(\bm{p}) \delta^3(\bm{p}-\bm{q}).
\label{eq:H0H1}
\end{equation}
Using them, we then find 
\begin{align}
\chi_t (0,a)
&=\beta_{t,t_0} \text{Tr}\Big[\int d^3p\,\hat{a}(\bm{p}) \rho(t_0)\hat{a}^\dagger(\bm{p})\Big]
+ \text{Tr}[ \big(\hat{\mathbb{I}}+(\gamma^*_{t,t_0 }+\gamma_{t,t_0}+|\gamma_{t,t_0}|^2)\hat{N} \big) e^{-ia_\mu \hat{P}^\mu }\rho(t_0)]
\nonumber 
\\
&+
 \text{Tr}\Big[ \int d^3p  \, \delta_{t,t_0} (p,p)
\hat{a}^\dagger (\bm{p})\hat{a}(\bm{p}) 
\, e^{-ia_\mu \hat{P}^\mu }\rho(t_0) \Big]
\nonumber 
\\
&=\beta_{t,t_0} \text{Tr}\big[\hat{N}  \rho(t_0)\big]
+ \text{Tr}[ \big(\hat{\mathbb{I}}+(\gamma^*_{t,t_0 }+\gamma_{t,t_0}+|\gamma_{t,t_0}|^2)\hat{N} \big) e^{-ia_\mu \hat{P}^\mu }\rho(t_0)]
+
\delta_{t,t_0} \text{Tr}\big[ \hat{N} 
\, e^{-ia_\mu \hat{P}^\mu }\rho(t_0) \big]
\nonumber 
\\
&=\beta_{t,t_0} \text{Tr}\big[\hat{N}  \rho(t_0)\big]
+ \text{Tr}[ \big(\hat{\mathbb{I}}-(\beta_{t,t_0}+\delta_{t,t_0})\hat{N} \big) e^{-ia_\mu \hat{P}^\mu }\rho(t_0)]
+
\delta_{t,t_0} \text{Tr}\big[ \hat{N} 
\, e^{-ia_\mu \hat{P}^\mu }\rho(t_0) \big]
\nonumber 
\\
&=
\chi_s (0,a)+
\beta_{t,t_0} \text{Tr}\big[\hat{N}\big(\hat{\mathbb{I}}-e^{-ia_\mu \hat{P}^\mu } \big) \rho(t_0)\big],
\nonumber
\end{align}
where in the second equality note that
$\delta_{t,t_0}(p,p)=\delta_{t,t_0}$ is independent of the three-momentum 
$\bm{p}$ due to the Lorentz invariance,
$\delta_{t,t_0}(p,p)=\delta_{t,t_0}(\Lambda p,\Lambda p)$, and in the third equality we used 
$\gamma^*_{t,t_0 }+\gamma_{t,t_0}+|\gamma_{t,t_0}|^2=-\beta_{t,t_0}-\delta_{t,t_0}$ satisfied by the second equation in \eqref{eq:comp0}. 
Hence, we get Eq.\eqref{eq:chit2}.

Let us compute the characteristic function 
$\chi_t(\theta,0)$. 
Using 
$e^{\frac{i}{2}\theta_{\mu\nu} \hat{J}^{\mu\nu}(t)}=e^{-i\hat{H}(t-t_0)}e^{\frac{i}{2}\theta_{\mu\nu} \hat{J}^{\mu\nu}(t_0)}e^{i\hat{H}(t-t_0)}$, which follows by 
$\hat{\bm{K}}(t)=e^{-i\hat{H}t} \hat{\bm{K}}(0)e^{i\hat{H}t}$ and 
$\hat{\bm
{J}}=e^{-i\hat{H}t} \hat{\bm{J}} e^{i\hat{H}t}$ (see also the discussion around Eq.\eqref{eq:KK_0}), we have
\begin{equation}
\chi_t (\theta,0)
=\text{Tr}[e^{\frac{i}{2} \theta_{\mu\nu} \hat{J}^{\mu\nu} (t)} \rho(t)]
=\text{Tr}[e^{\frac{i}{2} \theta_{\mu\nu} \hat{J}^{\mu\nu} (t)} \mathcal{U}_{t,t_0} \circ \mathcal{E}_{t,t_0} [\rho(t_0)]]
=\text{Tr}[e^{\frac{i}{2} \theta_{\mu\nu} \hat{J}^{\mu\nu} (t_0)}  \mathcal{E}_{t,t_0} [\rho(t_0)]]. 
\nonumber
\end{equation}
Substituting 
$\mathcal{E}_{t,t_0}$ into the above equation and assuming 
$\beta_{t,t_0}=0$, we obtain 
\begin{align}
\chi_t (\theta,0)
&= \text{Tr}[e^{\frac{i}{2} \theta_{\mu\nu} \hat{J}^{\mu\nu} (t_0)}(\hat{\mathbb{I}}+\gamma_{t,t_0} \hat{N})\rho(t_0)(\hat{\mathbb{I}}+\gamma^*_{t,t_0 }\hat{N})]
\nonumber 
\\
&
+
\text{Tr}\Big[e^{\frac{i}{2} \theta_{\mu\nu} \hat{J}^{\mu\nu} (t_0)} \int d^3p \int d^3q \, \delta_{t,t_0} (p,q)
\hat{a}^\dagger(\bm{p})\hat{a}(\bm{p}) 
\, \rho(t_0) \, 
\hat{a}^\dagger (\bm{q})\hat{a}(\bm{q}) \Big].
\nonumber
\end{align}
With the help of the invariance of the number operator 
$e^{\frac{i}{2} \theta_{\mu\nu} \hat{J}^{\mu\nu} (t_0)}\hat{N}e^{-\frac{i}{2} \theta_{\mu\nu} \hat{J}^{\mu\nu} (t_0)}=\hat{N}$ and the transformation rule 
$e^{\frac{i}{2} \theta_{\mu\nu} \hat{J}^{\mu\nu} (t_0)}\hat{a}^\dagger(\bm{p})\hat{a}(\bm{p})e^{-\frac{i}{2} \theta_{\mu\nu} \hat{J}^{\mu\nu} (t_0)}=\frac{E_{\bm{p}_\Lambda}}{E_{\bm{p}}} \hat{a}^\dagger(\bm{p}_\Lambda)\hat{a}(\bm{p}_\Lambda)$, the characteristic function 
$\chi_{t} (\theta,0)$ is computed as 
\begin{align}
\chi_t (\theta,0)
&= \text{Tr}[(\hat{\mathbb{I}}+\gamma_{t,t_0} \hat{N})e^{\frac{i}{2} \theta_{\mu\nu} \hat{J}^{\mu\nu} (t_0)}\rho(t_0)(\hat{\mathbb{I}}+\gamma^*_{t,t_0 }\hat{N})]
\nonumber 
\\
&+
 \text{Tr}\Big[ \int d^3p \int d^3q \, \delta_{t,t_0} (p,q)
\frac{E_{\bm{p}_\Lambda}}{E_{\bm{p}}} \hat{a}^\dagger(\bm{p}_\Lambda)\hat{a}(\bm{p}_\Lambda) 
\, e^{\frac{i}{2} \theta_{\mu\nu} \hat{J}^{\mu\nu} (t_0)}\rho(t_0) \, 
\hat{a}^\dagger (\bm{q})\hat{a}(\bm{q}) \Big]
\nonumber
\\
&= \text{Tr}[(\hat{\mathbb{I}}+\gamma^*_{t,t_0 }\hat{N})(\hat{\mathbb{I}}+\gamma_{t,t_0} \hat{N})e^{\frac{i}{2} \theta_{\mu\nu} \hat{J}^{\mu\nu} (t_0)}\rho(t_0)]
\nonumber 
\\
&+
 \text{Tr}\Big[ \int d^3p \int d^3q \, \delta_{t,t_0} (p,q)
\frac{E_{\bm{p}_\Lambda}}{E_{\bm{p}}} \hat{a}^\dagger (\bm{q})\hat{a}(\bm{q})\hat{a}^\dagger(\bm{p}_\Lambda)\hat{a}(\bm{p}_\Lambda) 
\, e^{\frac{i}{2} \theta_{\mu\nu} \hat{J}^{\mu\nu} (t_0)}\rho(t_0) 
 \Big]
 \nonumber
\\
&= \text{Tr}[\big(\hat{\mathbb{I}}+(\gamma^*_{t,t_0 }+\gamma_{t,t_0}+|\gamma_{t,t_0}|^2)\hat{N} \big)
e^{\frac{i}{2} \theta_{\mu\nu} \hat{J}^{\mu\nu} (t_0)}\rho(t_0)]
\nonumber 
\\
&+
 \text{Tr}\Big[ \int d^3p \, \delta_{t,t_0} (p,\Lambda p)
\frac{E_{\bm{p}_\Lambda}}{E_{\bm{p}}} \hat{a}^\dagger (\bm{p}_\Lambda)\hat{a}(\bm{p}_\Lambda) 
\, e^{\frac{i}{2} \theta_{\mu\nu} \hat{J}^{\mu\nu} (t_0)}\rho(t_0) 
 \Big]
\nonumber 
\\
&= \text{Tr}[\big(\hat{\mathbb{I}}-\delta_{t,t_0}\hat{N} \big)
e^{\frac{i}{2} \theta_{\mu\nu} \hat{J}^{\mu\nu} (t_0)}\rho(t_0)]
+
 \text{Tr}\Big[ \int d^3p \, \delta_{t,t_0} (p,\Lambda p)
\frac{E_{\bm{p}_\Lambda}}{E_{\bm{p}}} \hat{a}^\dagger (\bm{p}_\Lambda)\hat{a}(\bm{p}_\Lambda) 
\, e^{\frac{i}{2} \theta_{\mu\nu} \hat{J}^{\mu\nu} (t_0)}\rho(t_0) 
 \Big]
\nonumber 
\\
&= \chi_{t_0} (\theta,0)-\delta_{t,t_0}\text{Tr}[\hat{N} e^{\frac{i}{2} \theta_{\mu\nu} \hat{J}^{\mu\nu} (t_0)}\rho(t_0)]
+
 \text{Tr}\Big[ \int d^3p \, \delta_{t,t_0} (p, \Lambda p)
\hat{a}^\dagger (\bm{p})\hat{a}(\bm{p}) 
\, e^{\frac{i}{2} \theta_{\mu\nu} \hat{J}^{\mu\nu} (t_0)}\rho(t_0) 
 \Big],
 \nonumber 
\end{align}
where the second and the third equations of \eqref{eq:H0H1} were used in the third equality, and in the forth equality the equation 
$\gamma^*_{t,t_0 }+\gamma_{t,t_0}+|\gamma_{t,t_0}|^2=-\delta_{t,t_0}$ was substituted. 
The Lorentz invariace of 
$d^3p/E_{\bm{p}}$ and 
$\delta_{t,t_0}(p,q)$ leads to the last equality, which is nothing but Eq.\eqref{eq:chit3}.

\end{appendix}


\end{document}